\documentclass[12pt]{article}
\pdfoutput=1
\usepackage{indentfirst}
\usepackage{yfonts}
\usepackage{color}
\usepackage{mhchem}
\usepackage{xcolor}
\usepackage{harpoon}
\usepackage{cite}
\usepackage{hyperref}
\hypersetup{colorlinks=true,linkcolor=red,anchorcolor=black,citecolor=green}
\usepackage[toc,page]{appendix}
\usepackage{amsfonts}
\usepackage{bbold}
\usepackage{textcomp}
\usepackage[DIV13]{typearea}
\usepackage{amsmath, amsthm, amssymb, mathtools,empheq,latexsym,dsfont}
\usepackage{bbm}
\usepackage{slashed, simplewick}
\usepackage[utf8]{inputenc}
\usepackage{graphicx,placeins}
\usepackage{makeidx}
\usepackage[font=normalsize,labelfont=bf]{caption}
\usepackage{nicefrac}
\usepackage{subfig}
\usepackage{overpic}
\usepackage{array, bigdelim,multirow,multicol}
\usepackage[integrals]{wasysym}
\usepackage{fancybox}
\usepackage{bm}
\usepackage{float}
\usepackage{rotating}
\usepackage{colortbl}
\usepackage{booktabs}
\usepackage[top=2cm,textwidth=16.6cm,textheight=22.75cm]{geometry}
\usepackage{doi}
\usepackage{extarrows}
\usepackage{pifont}
\usepackage{mathrsfs}
\usepackage{lmodern}

\usepackage{braket}

\usepackage{tikz}

\usetikzlibrary{quotes,angles}
\graphicspath{{immagini/}}
\usetikzlibrary{arrows.meta}
\usetikzlibrary{decorations}
\usepackage{tikz-3dplot}
\usepackage{tikz-feynman}
\usepackage{mathrsfs}
\usepackage{enumerate}
\usetikzlibrary {arrows.meta}
\newcommand{\ignore}[1]{}
\usepackage{framed}
\usepackage{longtable}
\usepackage{makecell}
\usepackage{arydshln}
\usepackage{tabularx}
\usepackage{makecell}
\usepackage{diagbox}
\allowdisplaybreaks  
\graphicspath{{0nbb/}}

\definecolor{Gray}{gray}{0.92}

\newcommand{\be}{\begin{equation}}
\newcommand{\ee}{\end{equation}}
\newcommand{\bea}{\begin{eqnarray}}
\newcommand{\eea}{\end{eqnarray}}

\makeatletter
\renewcommand*{\@fnsymbol}[1]{\ensuremath{\ifcase#1\or *\or  \mathsection\or \ddagger\or
		\dagger\or \mathparagraph\or \|\or **\or \dagger\dagger
		\or \ddagger\ddagger \else\@ctrerr\fi}}
\makeatother

\begin{document}

\title{\Large\bf Systematical decomposition of dimension-11 short-range neutrinoless double beta decay operators}
\date{}
\author{Shi-Yu Li$^{a}$\footnote{E-mail: {\tt
lishiyu@mail.ustc.edu.cn}} \,,
Gui-Jun Ding$^{a,b}$\footnote{E-mail: {\tt
dinggj@ustc.edu.cn}}  \
\\*[20pt]
\centerline{
\begin{minipage}{\linewidth}
\begin{center}
$^a${\it \small Department of Modern Physics,  and Anhui Center for fundamental sciences in theoretical physics,\\
University of Science and Technology of China, Hefei, Anhui 230026, China}\\[2mm]
$^b${\it \small  College of Physics, Guizhou University, Guiyang 550025, China}\\[2mm]
\end{center}
\end{minipage}}
\\[10mm]}

\maketitle

\begin{abstract}

Neutrinoless double beta decay ($0\nu\beta\beta$) may receive sizable contributions from short-range physics beyond the Standard Model. We present a systematical classification of all tree-level ultraviolet completions of the dimension-11 short-range $0\nu\beta\beta$ decay operators, renormalizable scenarios with scalar and fermion mediators are considered. We identify eight distinct topologies and twenty-eight viable diagrams, from which all consistent UV completions are generated by imposing Standard Model gauge invariance. All these models involve a total of 61 new fields beyond the Standard Model and they typically feature fractionally charged fermions and exotic bosons such as dileptons, diquarks, and leptoquarks. We further study a representative model without colored mediators and analyze its implications for the $0\nu\beta\beta$ decay half-life and light neutrino masses. We find that current and future $0\nu\beta\beta$ decay experiments impose stringent constraints. Our systematic decomposition provides a general framework for exploring exotic short-range contributions to $0\nu\beta\beta$ decay in future experiments.

\end{abstract}
\thispagestyle{empty}
\vfill

\newpage

{\hypersetup{linkcolor=black}
\tableofcontents
}

\section{Introduction}

The discovery of neutrino oscillations implies that at least two of the three neutrinos have non-zero masses~\cite{McDonald:2016ixn,Kajita:2016cak}.  Neutrino oscillation experiments can measure the neutrino mass-squared differences $\Delta m^2_{ij}\equiv m^2_i-m^2_j\, (ij=21, 31, 32)$ as
$\Delta m^2_{21}=\left(7.236\rightarrow7.823\right)\times 10^{-5}\,\text{eV}^2$, $\Delta m^2_{31}=\left(2.450\rightarrow2.576\right)\times 10^{-3}\,\text{eV}^2$ for normal ordering spectrum and $\Delta m^2_{32}=\left(-2.547\rightarrow-2.421\right)\times 10^{-3}\,\text{eV}^2$ for inverted ordering spectrum at $3\sigma$ level~\cite{Esteban:2024eli}. Notice that only two of the three mass-squared differences are independent because of $\Delta m^2_{32}=\Delta m^2_{31}-\Delta m^2_{21}$. The absolute neutrino mass  rather than mass differences, can be extracted from energy-momentum conservation relation in reactions in which a neutrino or an antineutrino is involved. The most recent result on the kinematic search for neutrino mass in tritium decay $^3\text{H}\rightarrow ^3\text{He}+e^{-}+\bar{\nu}_e$ is from KATRIN which sets an upper limit on the effective neutrino mass $m_{\beta}\equiv\sqrt{ \sum_i |U_{ei}|^2m^2_i}<0.45\,\text{eV}$ at $90\%$ CL~\cite{KATRIN:2024cdt}, where $U_{ei}$ are the elements of the lepton mixing matrix $U$ and $m_i$ denote the light neutrino mass. In addition, cosmological observations from large scale structure, including the CMB, the distribution of clusters of galaxies,
and the Lyman-alpha forest, set an upper limit on the sum of neutrino masses $\sum m_{\nu}<0.12  \text{eV}$~\cite{Planck:2018vyg}.

There are only left-handed neutrinos in the standard model (SM) so that neutrinos are massless in SM. In order to generate tiny neutrino mass, one has to introduce new degree of freedom which can be one or several new fermions or bosons. For instance, neutrinos could be of Dirac particles as the other SM charged fermions by adding right-handed neutrinos to the SM particle content, nevertheless this requires tiny Yukawa couplings of order $10^{-12}$. The mass eigenstate of neutrino could be identical with that of antineutrino, i.e. neutrinos are Majorana particle. Taking the particle content of the minimal SM, there is a unique dimension-5 operator which is the famous Weinberg operator $\mathcal{O}_{W}=\frac{1}{\Lambda_{\text{NP}}}(\overline{L^c}i\sigma^2H)(H^{T}i\sigma^2L)$, where $L$ and $H$ are the SM lepton doublets and Higgs fields respectively. After the electroweak symmetry breaking, the Weinberg operator would lead to Majorana neutrino mass of order $v^2/\Lambda_{\text{NP}}$. The non-renormalizable Weinberg operator can be mediated by the exchange of a SM singlet fermion, a triplet scalar or a triplet fermion which correspond to the type I~\cite{Minkowski:1977sc,Yanagida:1979as,Gell-Mann:1979vob,Mohapatra:1979ia}, type II~\cite{Magg:1980ut,Schechter:1980gr,Wetterich:1981bx,Lazarides:1980nt,Mohapatra:1980yp,Cheng:1980qt} and type III~\cite{Foot:1988aq} seesaw mechanism respectively.

If neutrinos are Majorana fermions, the lepton number would be violated by two units by the neutrino Majorana mass term. If neutrinos are Dirac particles, lepton number conservation is an exact law of nature and consequently the $\Delta L=2$ processes would be forbidden. It is known that the search for neutrinoless double beta decay ($0\nu\beta\beta$ decay) is are the most sensitive probe of lepton number violation and Majorana neutrino mass. $0\nu\beta\beta$ decay is a $\Delta L=2$ rare nuclear decay process of the type $(Z, A)\rightarrow (Z+2, A)+2e^{-}$ where the atomic number $Z$ increases by two units while the nucleon number $A$ remains constant. Hence this process takes place in a nucleus and converts two neutrons
into two protons and two electrons without outgoing antineutrinos. A positive signal of $0\nu\beta\beta$ decay would indicates that neutrinos are Majorana particles~\cite{Schechter:1981bd}. Moreover, the observation of $0\nu\beta\beta$ can provide important clues for understanding the neutrino mass generation mechanism and the origin of the matter-antimatter asymmetry in our Universe. In addition, the $0\nu\beta\beta$ can also provide insight to the flavor structure of SM, and it is a powerful tool to test and discriminate flavor models. See Refs.~\cite{Dolinski:2019nrj,Ejiri:2019ezh,Agostini:2022zub,Cirigliano:2022oqy} for recent review on $0\nu\beta\beta$ decay.

There are plenty of experiments searching for $0\nu\beta\beta$ decay with different isotopes. The most stringent limit on the $^{136}\textrm{Xe}$ decay half-life is  $T^{0\nu}(^{136}\textrm{Xe})>3.8\times 10^{26}$ yr from KamLAND-Zen~\cite{KamLAND-Zen:2024eml} at $90\%$ confidence level, and the effective Majorana neutrino mass is constrained to be in the region $|m_{\beta\beta}|\equiv |\sum_i U^2_{ei}m_i|<(28-122)$ meV where the uncertainty on this bound arises from the nuclear matrix element. The most sensitive limit on the $^{76}\text{Ge}$ $0\nu\beta\beta$ decay half-life is $T_{1/2}(^{76}\text{Ge})>1.8\times10^{26}$ yr from GERDA~\cite{GERDA:2020xhi}, which implies a bound $|m_{\beta\beta}|<(79-180)$ meV. The next generation $0\nu\beta\beta$ experiments will increase the half-life sensitivity by a factor of about 100,  thus the effective Majorana neutrino mass $|m_{\beta\beta}|$ can be probed down to about 10 meV.

It is usually assumed that the $0\nu\beta\beta$ decay is dominantly induced by the exchange of light Majorana neutrinos between two charged current interaction vertices, and thus the decay amplitude is proportional to the effective mass $m_{\beta\beta}$. Consequently it is usually called mass mechanism. A priori one doesn't know whether the mass mechanism dominates the $0\nu\beta\beta$ decay. The possible physics beyond the SM can also contribute to the $0\nu\beta\beta$ decay, if it contains lepton number violating interactions. The energy transfer in $0\nu\beta\beta$ decay is typically a few MeV. Thus at low energy below the electroweak scale, the SM gauge symmetry $SU(3)_C\times SU(2)_L\times U(1)_Y$ is spontaneously broken down to $SU(3)_C\times U(1)_{EM}$ and the most general Lagrangian responsible for $0\nu\beta\beta$ decay consists of the following 24 independent terms~\cite{Pas:2000vn}, and they can be written as the product of three currents of quark and lepton fields as follows,
\begin{equation}
\label{eq:Leff-short}
{\cal L}_{\rm eff} = \frac{G_F^2}{2 m_p} \left\{ \epsilon_1^X JJj + \epsilon_2^X J^{\mu\nu} J_{\mu \nu} j + \epsilon_3^X J^\mu J_\mu j + \epsilon_4^X J^\mu J_{\mu\nu} j^\nu +
\epsilon_5^X J^\mu J j_\mu
\right\}\,,
\end{equation}
where $J = \bar u (1\pm\gamma_5)d$, $j = \bar e (1 \pm \gamma_5) e^c$,
$J_{\mu\nu} = \bar u \frac i 2 [\gamma_\mu, \gamma_\nu] (1 \pm
\gamma_5)d$, $J_\mu = \bar u \gamma_\mu (1 \pm \gamma_5) d$, and
$j_\mu = \bar e \gamma_\mu (1 \pm \gamma_5) e^c$ are bilinear currents of quarks or leptons. The chirality of the operators is encoded in $X = abc$, where $a,b,c$ is $L$ or $R$ and they refer to the chirality of the three fermion currents respectively. Pinning down the origin of the $0\nu\beta\beta$ decay requires unraveling the nature of the UV completions responsible for the effective operators in Eq.~\eqref{eq:Leff-short}. If the $0\nu\beta\beta$ decay is induced by the exchange of a light neutrino between two nucleons at low energy, the effective Lagrangian would be the four-fermion interactions~\cite{Pas:1999fc,Deppisch:2012nb} and it is called long-range mechanism. The decomposition of long-range effective operators has been studied at tree~\cite{Helo:2016vsi} and one-loop level~\cite{Chen:2023xaz}. In the present work, we shall focus on the $0\nu\beta\beta$ decay  mediated only by heavy mediators with masses above the electroweak scale, it is called short-range mechanism in the literature~\cite{Pas:2000vn,Deppisch:2012nb}. Integrating out the heavy mediators, the $0\nu\beta\beta$ decay can be described by non-renormalizable operators which are made of the SM fields and invariant under the SM gauge group~\cite{Pas:1999fc,Pas:2000vn,Babu:2001ex,deGouvea:2007qla,delAguila:2012nu}. These operators involve at least four quark fields and two lepton fields, consequently the lowest dimension is equal to nine. It turns out that there are eleven electroweak invariant operators at dimension-9 and they lead to eleven of the twenty-four operators in Eq.~\eqref{eq:Leff-short} at low energy~\cite{Graesser:2016bpz}. The other thirteen operators in Eq.~\eqref{eq:Leff-short} don't occur at dimension-9 because they violate SM gauge invariance, twelve of them are found to appear at dimension-11 and one of them at dimension-13. Systematical analysis of the ultraviolet (UV) completions of the dimension-9 $0\nu\beta\beta$ operators was presented in Refs.~\cite{Bonnet:2012kh,Chen:2021rcv}, it was found that there are two and six different topologies at tree~\cite{Bonnet:2012kh} and one-loop~\cite{Chen:2021rcv} level respectively.

In this work, we will focus on the $0\nu\beta\beta$ operators at dimension-11 that after electroweak symmetry reduce to those operators in Eq.~\eqref{eq:Leff-short} which have not previously been found at dimension-9. The aim is to identify all possible tree-level decompositions of these effective operators and the corresponding new messenger fields would be listed out. We require that the messengers can not mediate the short-range $0\nu\beta\beta$ operators of dimension-9, otherwise the contributions of the concerned dimension-11 operators would be just some minor corrections to the $0\nu\beta\beta$ decay. Due to the presence of the
$\Delta L=2$ interactions in the decomposition of the $0\nu\beta\beta$ operators, the light neutrinos are Majorana particles and the neutrino masses mediated by the $0\nu\beta\beta$ messenger fields can be generated.
The UV completions of these $0\nu\beta\beta$ decay operators typically involve fractionally charged fermions and exotic bosons such as dileptons, diquarks and leptoquarks, leading to rich collider and low-energy phenomenology. Our analysis offers a general framework for probing exotic short-range contributions to $0\nu\beta\beta$ decay in future experiments.

The rest of this paper is organized as follows. We shall present the short-range $0\nu\beta\beta$ decay operators at dimension-11 in section~\ref{sec:0nbb-optrs-dim11}. We provide all possible topologies, diagrams and models together with the SM quantum numbers of the mediators in section~\ref{sec:top-diag-models-UV}. The most minimal example model at dimension-11 is given in section~\ref{sec:representative-model}, and the predictions for the half-life of $0\nu\beta\beta$ decay and the light neutrino mass are studied. We summarize our results and present our conclusions in section~\ref{sec:conclusion}. The tree-level non-renormalizable topologies for the dim-11 $0\nu\beta\beta$ decay operators are collected in appendix~\ref{app:non-renor-top}. The master formula for the short-range $0\nu\beta\beta$ decay half-life, together with the corresponding low-energy effective operators, is given in appendix~\ref{sec:0nubb-eff-operators-LEFT}. The explicit expressions for the two-loop integrals relevant for the evaluation of neutrino masses are provided in appendix~\ref{app:2loop-def}. Finally, in appendix~\ref{app:add-onbb-models} we present several additional simple models in addition to the representative example discussed in section~\ref{sec:representative-model}, focusing on scenarios with at most one color-triplet mediator and otherwise colorless fields.

\section{\label{sec:0nbb-optrs-dim11}Short-range $0\nu\beta\beta$ decay operators of dimension-11}

Twelve of the low energy $0\nu\beta\beta$ operators in Eq.~\eqref{eq:Leff-short} that don't appear at dimension-9 can be made $SU(3)_C\times SU(2)_L \times U(1)_Y$ gauge invariant through the insertion of two additional Higgs fields among the six quark and lepton fields. It was found that there are totally nineteen independent short-range $0\nu\beta\beta$ decay operators compatible with SM gauge symmetry at dimension $d=11$\cite{Graesser:2016bpz}. Here we have dropped the operators which are the SM gauge invariant dimension-9 $0\nu\beta\beta$ decay operators multiplied with $H^\dagger H$, since they can not generate low energy operators interesting to us. Considering the constituent fields of each operator, we can classify these operators into the following seven different types of operators~\cite{Graesser:2016bpz}:
\begin{eqnarray}
\nonumber \mathcal{O}_{1a}&=&\epsilon_{ab}H^*_aH^*_c\left(\overline{Q_L}_b\gamma^\mu {Q_L}_c\right)\left(\overline{u_R}\gamma_\mu d_R\right)\left(\overline{e_R}e_R^C\right)\,,\\
\nonumber \mathcal{O}_{1b}&=&\epsilon_{ab}H^*_aH^*_c\left(\overline{Q_L}_b\gamma^\mu\lambda^A{Q_L}_c\right)\left(\overline{u_R}\gamma_\mu\lambda^Ad_R\right)\left(\overline{e_R}e_R^C\right)\,, \\
\nonumber \mathcal{O}_{2a}&=&H^*_aH^*_b\left(\overline{u_R}{Q_L}_a\right)\left(\overline{u_R}{Q_L}_b\right)\left(\overline{e_R}e_R^C\right)\,, \\
\nonumber \mathcal{O}_{2b}&=&H^*_aH^*_b\left(\overline{u_R}\lambda^A{Q_L}_a\right)\left(\overline{u_R}\lambda^A{Q_L}_b\right)\left(\overline{e_R}e_R^C\right) \,, \\
\nonumber \mathcal{O}_{3a}&=&\epsilon_{ab}\epsilon_{cd}H^*_aH^*_c\left(\overline{{Q_L}_b}d_R\right)\left(\overline{Q_L}_dd_R\right)\left(\overline{e_R}e_R^C\right) \,, \\
\nonumber \mathcal{O}_{3b}&=&\epsilon_{ab}\epsilon_{cd}H^*_aH^*_c\left(\overline{Q_L}_b\lambda^Ad_R\right)\left(\overline{Q_L}_d\lambda^Ad_R\right)\left(\overline{e_R}e_R^C\right)\,,\\
\nonumber \mathcal{O}_{4}&=&H_aH_b\left(\overline{u_R}\gamma^\mu d_R\right)\left(\overline{u_R}\gamma_\mu d_R\right)\left(\overline{\ell_L}_a{\ell^C_L}_b\right)\,, \\
\nonumber \mathcal{O}_{5a}&=&\epsilon_{ae}\epsilon_{cf}H_b^*H_d^*\left(\overline{Q_L}_a\gamma^\mu {Q_L}_b\right)\left(\overline{Q_L}_c\gamma_\mu {Q_L}_d\right)\left(\overline{\ell_L}_e{\ell^C_L}_f\right)\,,\\
\nonumber \mathcal{O}_{5b}&=&\epsilon_{ae}\epsilon_{cf}H_e^*H_d^*\left(\overline{Q_L}_a\gamma^\mu {Q_L}_b\right)\left(\overline{Q_L}_c\gamma_\mu {Q_L}_d\right)\left(\overline{\ell_L}_b{\ell^C_L}_f\right)\,, \\
\nonumber\mathcal{O}_{5c}&=&\epsilon_{ae}\epsilon_{cf}H_e^*H_c^*\left(\overline{Q_L}_a\gamma^\mu {Q_L}_b\right)\left(\overline{Q_L}_f\gamma_\mu {Q_L}_d\right)\left(\overline{\ell_L}_b\ell^C_{Ld}\right)\,, \\
\nonumber \mathcal{O}_{5d} &=& \epsilon_{ae} \epsilon_{cf} H^*_{c} H^*_{d}
(\overline{Q_L}_{e} \gamma^\mu  {Q_L}_b)(\overline{Q_L}_{f} \gamma_\mu {Q_L}_d)(\overline{\ell_L}_ {a}  {\ell^C_L}_b) \,,\\
\nonumber\mathcal{O}_{6a}&=&\epsilon_{ae}\epsilon_{cd}H^*_eH^*_b\left(\overline{Q_L}_a\gamma^\mu {Q_L}_b\right)\left(\overline{Q_L}_cd_R\right)\left(\overline{\ell_L}_d\gamma_\mu e_R^C\right) \,, \\
\nonumber\mathcal{O}_{6b}&=&\epsilon_{ae}\epsilon_{cd}H^*_eH^*_b\left(\overline{{Q_L}_a}\gamma^\mu\lambda^A{Q_L}_b\right)\left(\overline{Q_L}_c\lambda^Ad_R\right)\left(\overline{\ell_L}_d\gamma_\mu e_R^C\right)\,, \\
\nonumber\mathcal{O}_{7a}&=&\epsilon_{ab}H^*_bH^*_c\left(\overline{Q_L}_a\gamma^\mu {Q_L}_c\right)\left(\overline{u_R}{Q_L}_d\right)\left(\overline{\ell_L}_d\gamma_\mu e_R^C\right)\,, \\
\nonumber\mathcal{O}_{7b}&=&\epsilon_{ab}H^*_bH^*_d\left(\overline{Q_L}_a\gamma^\mu {Q_L}_c\right)\left(\overline{u_R}{Q_L}_d\right)\left(\overline{\ell_L}_c\gamma_\mu e_R^C\right)\,,\\
\nonumber\mathcal{O}_{7c}&=&\epsilon_{ab}H^*_cH^*_d\left(\overline{Q_L}_a\gamma^\mu {Q_L}_c\right)\left(\overline{u_R}{Q_L}_d\right)\left(\overline{\ell_L}_b\gamma_\mu e_R^C\right)\,,\\
\nonumber\mathcal{O}_{7d}&=&\epsilon_{ab}H^*_bH^*_c\left(\overline{Q_L}_a\gamma^\mu\lambda^A {Q_L}_c\right)\left(\overline{u_R}\lambda^A{Q_L}_d\right)\left(\overline{\ell_L}_d\gamma_\mu e_R^C\right)\,,\\
\nonumber\mathcal{O}_{7e}&=&\epsilon_{ab}H^*_bH^*_d\left(\overline{Q_L}_a\gamma^\mu\lambda^A {Q_L}_c\right)\left(\overline{u_R}\lambda^A{Q_L}_d\right)\left(\overline{\ell_L}_c\gamma_\mu e_R^C\right)\,,\\
\mathcal{O}_{7f}&=&\epsilon_{ab}H^*_cH^*_d\left(\overline{Q_L}_a\gamma^\mu\lambda^A {Q_L}_c\right)\left(\overline{u_R}\lambda^A{Q_L}_d\right)\left(\overline{\ell_L}_b\gamma_\mu e_R^C\right)\,, \label{eq:0nbb-d11-opers}
\end{eqnarray}
where $Q_L=(u_L, ~d_L)^{T}$ and $\ell_L=(\nu_L,~e_L)^{T}$ refer to the quark and lepton doublets of $SU(2)_L$ respectively, $H$ is the SM Higgs doublet, and we denote the charge conjugation fields $e^c_R=(e_R)^c$ and $\ell_L^c =((\nu_L)^{c},~(e_L)^c)^T$. The Roman letters $i, j, m, n=1,2$ are the $SU(2)_L$ indices, and $\epsilon$ is the 2-component antisymmetric tensor with $\epsilon_{12}=-\epsilon_{21}=1$, and $\lambda^A$ are the Gell-Mann matrices. Notice that the $\mathcal{O}_{1a}$, $\mathcal{O}_{1b}$, $\mathcal{O}_{2a}$, $\mathcal{O}_{2b}$, $\mathcal{O}_{3a}$, $\mathcal{O}_{3b}$ involves the lepton fieds $\overline{e_R}e_R^C$.
$\mathcal{O}_{4}$, $\mathcal{O}_{5a}$, $\mathcal{O}_{5b}$, $\mathcal{O}_{5c}$, $\mathcal{O}_{5d}$  involve the lepton fields  $\overline{\ell_L}\ell_L^C$, and the remaining operators  $\mathcal{O}_{6a}$,  $\mathcal{O}_{6b}$, $\mathcal{O}_{7a}$,  $\mathcal{O}_{7b}$, $\mathcal{O}_{7c}$, $\mathcal{O}_{7d}$, $\mathcal{O}_{7e}$,  $\mathcal{O}_{7f}$ contain the lepton fields $\overline{\ell_L}\gamma^\mu e_R^C$. Here we only consider operators involving the first generation quark and lepton fields, since we are concerned with the operators contributing to the $0\nu\beta\beta$ decay. Notice that one can always multiply the SM invariant short-range $0\nu\beta\beta$ operators of dimension-9 by the combination $H^{\dagger} H$ to obtain dimension-11 operators. However, they reduce to the same low energy operators as those at dimension-9 and the corresponding contributions are suppressed by $v^2/\Lambda$, where $v$ refers to the Higgs vacuum expectation value and $\Lambda$ is the scale of lepton number violation.

At low energy, the SM gauge symmrtry $SU(3)_C\times SU(2)_L\times U(1)_Y$ is spontaneously broken into $SU(3)_C\times U(1)_{EM}$. Then the above operators leads to twelve short-range $0\nu\beta\beta$ operators which can be written as the product of three fermion currents~\cite{Pas:2000vn,Deppisch:2020ztt}, as summarized in table~\ref{tab:0nbb-operators-d11}. Notice that these low energy operators are different from those arising from the SM invariant dimension-9 operators~\cite{Chen:2021rcv}.

\begin{table}[htbp]
\centering
\begin{tabular}{|c|c|c|}\hline\hline
Operators& Operators after EWSB & Constituent fields  \\ \hline
$\mathcal{O}_{1a}$&$-\frac{v^2}{16}(\mathcal{O}_3)_{\{LR\}R}$&\multirow{2}{*}{$H^\dagger$, $H^\dagger$, $\overline{Q_L}$, $Q_L$, $\overline{u_R}$, $d_R$, $\overline{e_R}$, $e_R^C$}\\ \cline{1-2}
$\mathcal{O}_{1b}$&$\frac{v^2}{24}(\mathcal{O}_3)_{\{LR\}R}+\frac{v^2}{4}(\mathcal{O}_1)_{\{LR\}R}$&\\ \hline

$\mathcal{O}_{2a}$&$\frac{v^2}{16}(\mathcal{O}_1)_{\{LL\}R}$& \multirow{2}{*}{$H^\dagger$, $H^\dagger$, $\overline{u_R}$, $Q_L$, $\overline{u_R}$, $Q_L$, $\overline{e_R}$, $e_R^C$}\\ \cline{1-2}
$\mathcal{O}_{2b}$&$-\frac{5v^2}{48}(\mathcal{O}_1)_{\{LL\}R}-\frac{v^2}{64}(\mathcal{O}_2)_{\{LL\}R}$&\\\hline

$\mathcal{O}_{3a}$&$\frac{v^2}{16}(\mathcal{O}_1)_{\{RR\}R}$&\multirow{2}{*}{$H^\dagger$, $H^\dagger$, $\overline{Q_L}$, $d_R$, $\overline{Q_L}$, $d_R$, $\overline{e_R}$, $e_R^C$}\\ \cline{1-2}
$\mathcal{O}_{3b}$&$-\frac{5v^2}{48}(\mathcal{O}_1)_{\{RR\}R}-\frac{v^2}{64}(\mathcal{O}_2)_{\{RR\}R}$&\\ \hline

$\mathcal{O}_{4}$&$\frac{v^2}{16}(\mathcal{O}_3)_{\{RR\}L}$&$H$, $H$, $\overline{u_R}$, $d_R$, $\overline{u_R}$, $d_R$, $\overline{\ell_L}$, $\ell_L^C$\\ \hline

$\mathcal{O}_{5a}$, $\mathcal{O}_{5b}$, $\mathcal{O}_{5c}$, $\mathcal{O}_{5d}$ &$\frac{v^2}{16}(\mathcal{O}_3)_{\{LL\}L}$&$H^\dagger$, $H^\dagger$, $\overline{Q_L}$, $Q_L$, $\overline{Q_L}$, $Q_L$, $\overline{\ell_L}$, $\ell_L^C$\\ \hline
$\mathcal{O}_{6a}$&$-\frac{v^2}{16}(\mathcal{O}_5)_{RL}$&\multirow{2}{*}{$H^\dagger$, $H^\dagger$, $\overline{Q_L}$, $Q_L$, $\overline{Q_L}$, $d_R$, $\overline{\ell_L}$, $e_R^C$}\\ \cline{1-2}
$\mathcal{O}_{6b}$&$-\frac{5v^2}{48}(\mathcal{O}_5)_{RL}-\frac{iv^2}{16}(\mathcal{O}_4)_{LR}$&\\ \hline

$\mathcal{O}_{7a}$, $\mathcal{O}_{7b}$, $\mathcal{O}_{7c}$ & $-\frac{v^2}{16}(\mathcal{O}_5)_{LL}$&\multirow{2}{*}{$H^\dagger$, $H^\dagger$, $\overline{Q_L}$, $Q_L$,  $\overline{u_R}$, $Q_L$, $\overline{\ell_L}$, $e_R^C$}\\ \cline{1-2}

$\mathcal{O}_{7d}$, $\mathcal{O}_{7e}$, $\mathcal{O}_{7f}$&$\frac{5v^2}{48}(\mathcal{O}_5)_{LL}-\frac{iv^2}{16}(\mathcal{O}_4)_{LL}$&\\ \hline\hline
\end{tabular}
\caption{\label{tab:0nbb-operators-d11}The dim-11 $0\nu\beta\beta$ decay operators compatible with SM gauge symmetry and the corresponding contributions to low energy $0\nu\beta\beta$ decay operators. Here EWSB is the abbreviation of electroweak symmetry breaking, and $v$ is the vacuum expectation value of the Higgs field.    }
\end{table}

\section{\label{sec:top-diag-models-UV}UV completions of the dim-11 $0\nu\beta\beta$ decay operators}

In the present work, we are interested in the tree-level renormalizable UV completions of the dimension-11 $0\nu\beta\beta$ decay operators,  and we focus on the models in which the messengers are scalar or fermion fields. The cases meditated by the SM gauge bosons or exotic vectors\footnote{It is generally expected that the exotic vectors are gauge bosons of a new symmetry, thus one needs to extend the SM gauge group.} are neglected, although many of our results for scalar mediators also apply to diagrams with vector mediators. The new fermion fields are assumed to be vector-like fermions under the SM gauge symmetry in order to avoid the chiral anomaly. Following the diagrammatic approach developed in~\cite{Antusch:2008tz,Gavela:2008ra,Bonnet:2009ej,Bonnet:2012kh,Bonnet:2012kz,AristizabalSierra:2014wal,Cepedello:2018rfh}, we shall first find out the topologies of the tree-level Feynman diagrams for the concerned $0\nu\beta\beta$ operators, then specify the Lorentz nature (spinor or scalar) of each line, and finally fix the SM gauge quantum numbers of each messenger field by gauge invariance.

\subsection{Topologies}

\begin{figure}[hptb]
\centering
\includegraphics[scale=0.35]{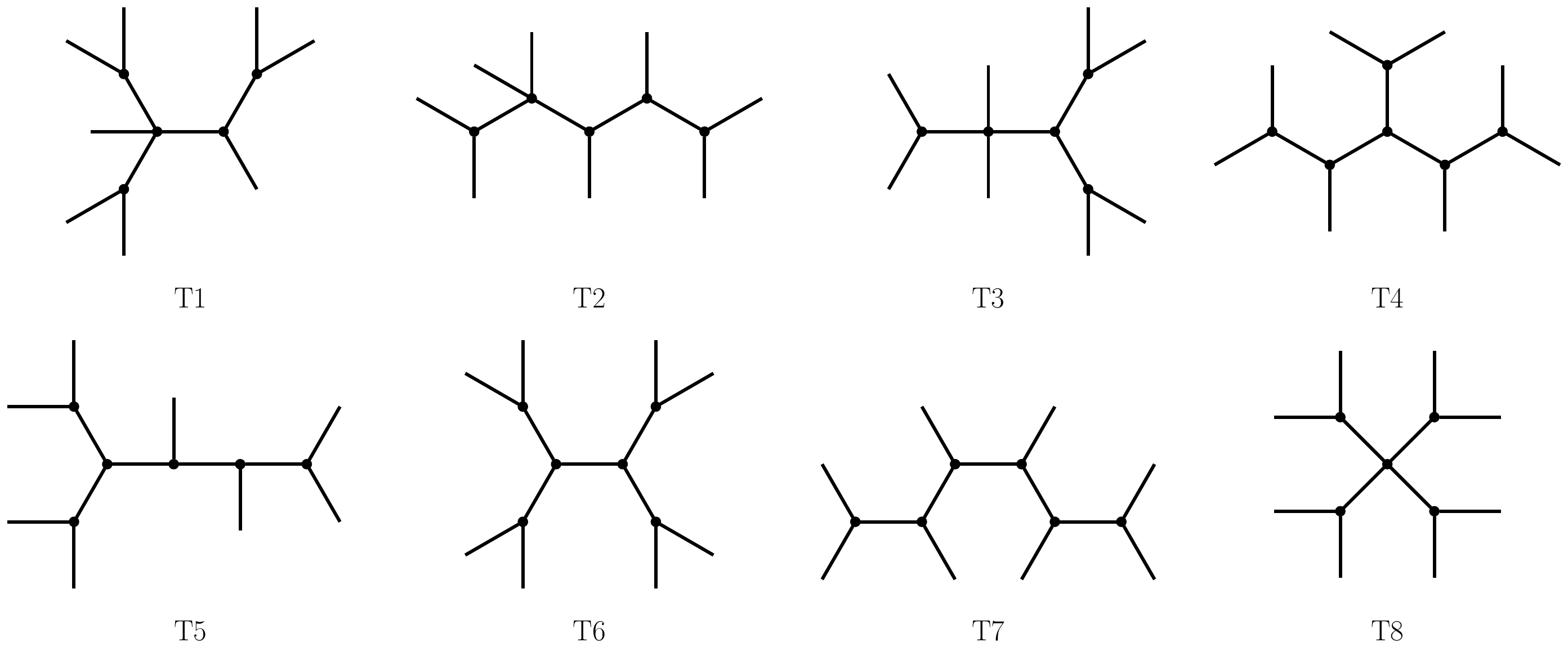}
\caption{The tree-level topologies for the UV completions of the $d=11$ $0\nu\beta\beta$ decay operators. The topologies requiring non-renormalizable interaction vertices are omitted here. }
\label{fig:topology}
\end{figure}

The dim-11 $0\nu\beta\beta$ decay operators in Eq.~\eqref{eq:0nbb-d11-opers}
involve eight fields including four quark fields, two lepton fields and two Higgs fields. We use our own \texttt{Mathematica} code to generate the topologies of the tree-level diagrams with eight external legs, where the Lorentz nature of neither internal or external fields are not specified. We find that there are totally eight different topologies relevant to these dimension-11 $0\nu\beta\beta$ operators, and they are displayed in figure~\ref{fig:topology}. Here we have dropped the eleven tree-level topologies in figure~\ref{fig:non-renormalizable-topologies}, since non-renormalizable interaction vertices are required in these diagrams.

We see that generally four mediators fields are necessary in the topologies T1, T2, T3 and T8 while five mediators are needed in the remaining topologies. The renormalizability limits the possible vertices to only three point scalar-scalar-scalar (SSS) interaction, fermion-fermion-scalar (FFS) interaction and four point scalar-scalar-scalar-scalar (SSSS) interaction.
As a consequence, the fields attached to a four point vertex have to be scalar fields. Thus all the four messenger fields in the last topology T8 are scalars. We see that there are four-point vertices involving external legs in the topologies T1, T2 and T3, the two external lines for Higgs fields are fixed for the topologies T2 and T3 while only one external Higgs line can be fixed for T1.

\subsection{Diagrams}

\begin{figure}[t!]
\centering
\includegraphics[scale=0.4]{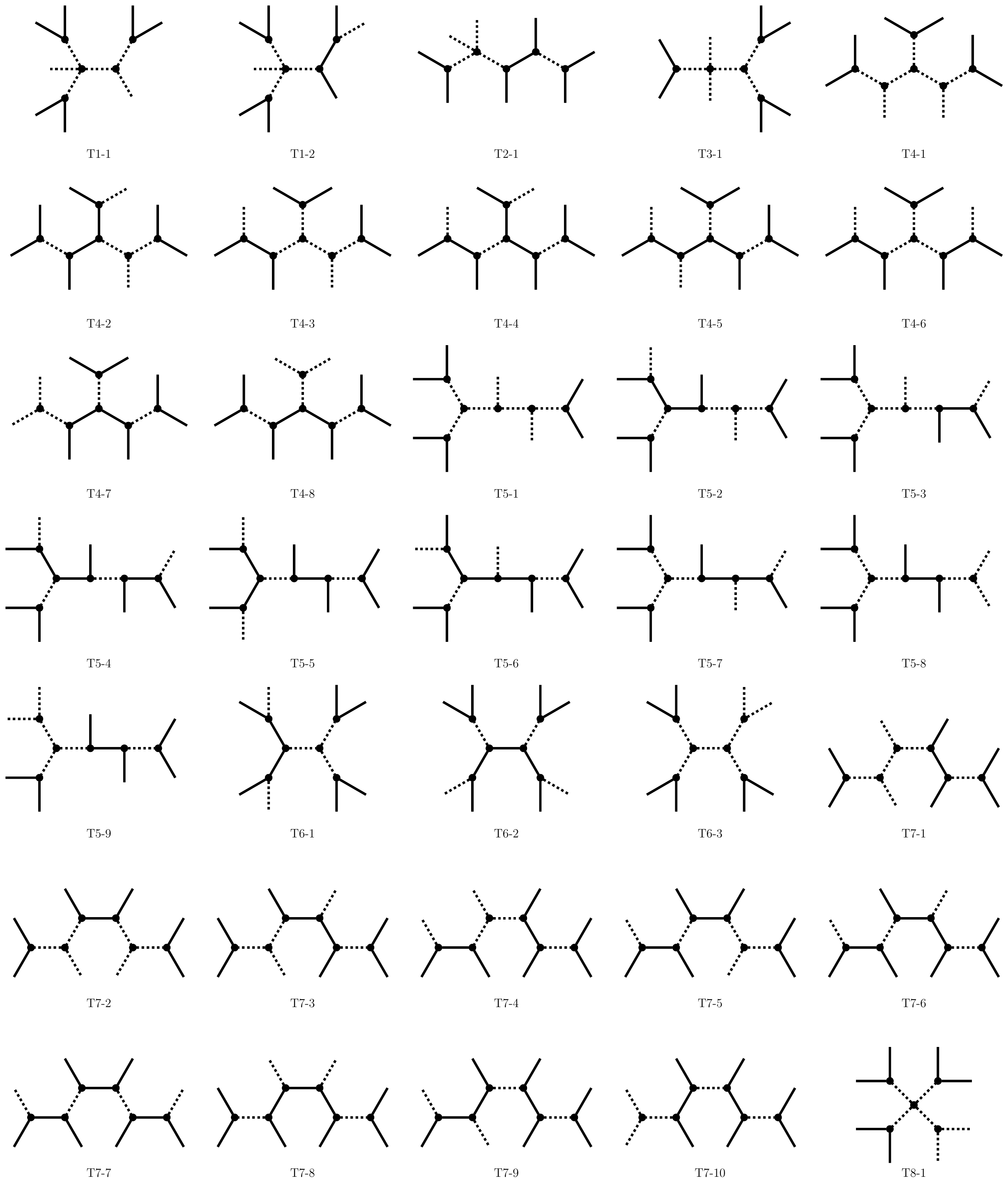}
\caption{\label{fig:genuine-diagram} All possible diagrams for the topologies in figure~\ref{fig:topology}, where the Lorentz nature (scalar and fermion) of each field is indicated by (dashed and solid) line.  }
\end{figure}

Then we proceed to promote the topologies in figure~\ref{fig:topology} into diagrams by specifying the Lorentz nature of each line, it can be either scalar or fermion field, as explained in above. The scalar and fermion fields are represented by dashed lines and solid lines respectively. The eight external lines invlove six fermions (four quark fields and two lepton fields) and two scalars (Higgs field). We firstly assign the fermion or scalar fields to the external fields. It is convenient to determine which two out of the eight external lines are Higgs fields, and then the remaining ones are fermion fields. It is notable that all the lines attached to a four point vertex are scalar fields, as explained in above. Hence there is only one diagram for each of the topologies T2, T3 and T8, as shown in figure~\ref{fig:genuine-diagram}. Moreover, Lorentz invariance and renormalizability require that each vertex can involve at most two fermion fields. As a consequence, each diagram has three fermion chains connecting the six external quark and lepton fields, and any two fermion chains can not cross each other, since the four fermion interactions are non-renormalizable. Usually several diagrams can be constructed for a given topology, some of them are identical due to the symmetry of the topology, and all the redundant diagrams should be removed. For instance, two independent diagrams T1-1 and T1-2 can be constructed from the first topology T1,  as shown in figure~\ref{fig:genuine-diagram}. One of the Higgs field is attached to the four-point vertex on the left and another Higgs field can be attached to either of two vertices on the right. In a similar way, we can find out the possible diagrams for each topology. The resulting tree level diagrams for two external scalars and six external fermions are displayed in figure~\ref{fig:genuine-diagram}, there are totally 35 possible diagrams.

\begin{figure}[htbp]
\includegraphics[scale=0.36]{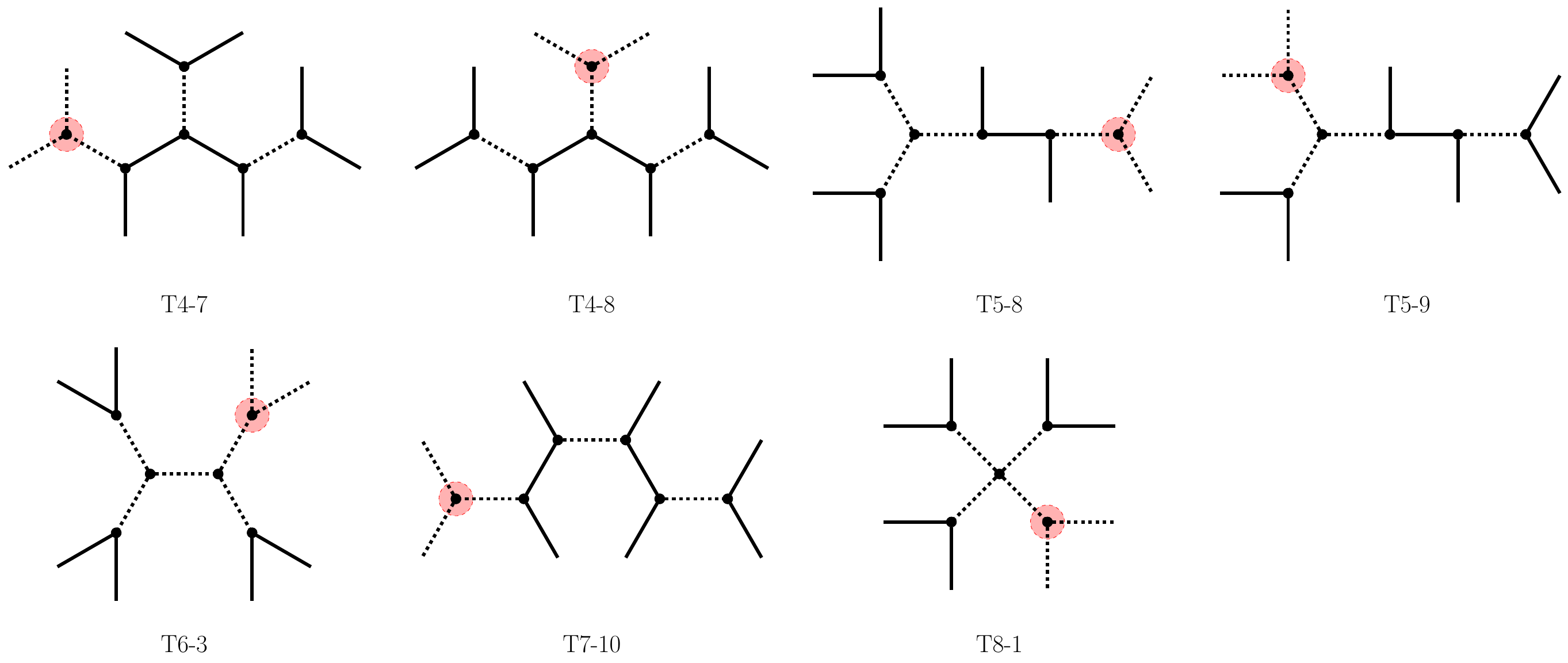}
\caption{\label{fig:type-II-seesaw-nongenuine}The $0\nu\beta\beta$ diagrams in which the two external Higgs bosons are attached to a single 3-point vertex.  }
\end{figure}

\begin{figure}[htbp]
\centering
\includegraphics[scale=0.5]{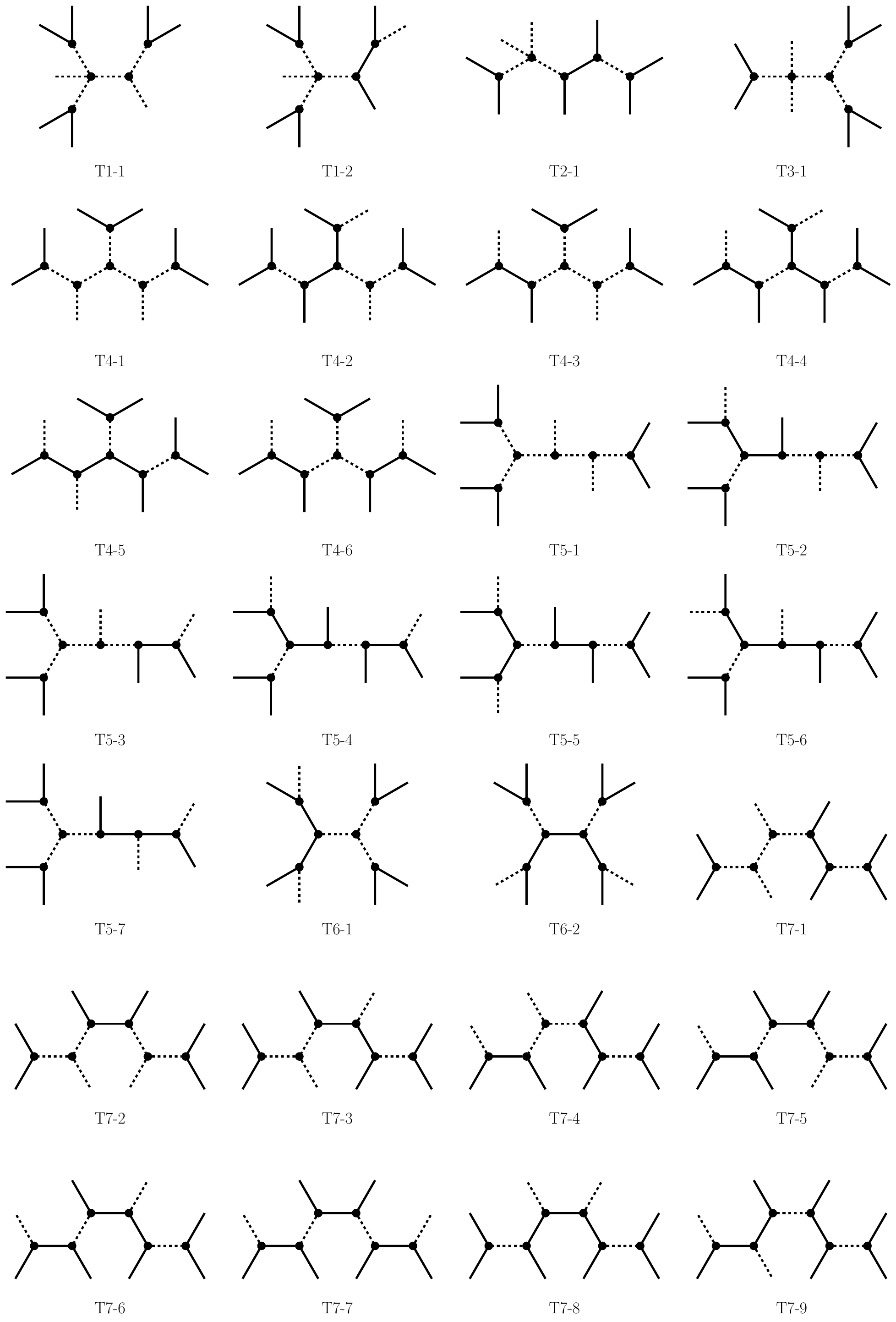}
\caption{\label{fig:genuine-diagrams}The tree-level diagrams for the UV completion of the dim-11 $0\nu\beta\beta$ decay effective operators, we here dropped the cases in which the two external Higg fields are attached to a 3-point vertex. }
\end{figure}

\subsection{Models}

As a next step, we assign the quark fields, lepton fields and Higgs fields involved in the dim-11 $0\nu\beta\beta$ operators to the external legs. Once a particular assignment is chosen, the quantum numbers of the internal fields under the SM gauge symmetry $SU(3)_C\times SU(2)_L\times U(1)_Y$ can be determined from the SM gauge invariance of each vertex. In the following, the SM gauge charges and Lorentz nature of internal fields are denoted as $(\bm{n}_3, \bm{n}_2, Y, \mathcal{L})$, where $\bm{n}_3$ and $\bm{n}_2$ stand for the irreducible representations of $SU(3)_C$ and $SU(2)_L$ respectively, $Y$ refers to the hypercharge, and $\mathcal{L}$ denotes the Lorentz nature, i.e., scalar (S) or fermion (F). In this notation, the SM gauge charges of the left-handed quark $Q_{L}$ and the Higgs field $H$ are $(\bm{3}, \bm{2}, \frac{1}{6}, F)$ and $(\bm{1}, \bm{2}, \frac{1}{2}, S)$ respectively. For any given diagrams in figure~\ref{fig:genuine-diagram} and any $0\nu\beta\beta$ operators in Eq.~\eqref{eq:0nbb-d11-opers}, there are generally multiple ways to assign relevant fields to external legs.

In the present work, we will focus on the dim-11 $0\nu\beta\beta$ decay models at tree-level. Our aim is to give a systematical analysis of such models, identify the possible tree-level topologies and diagrams. Thus the tree level contributions from the dim-9 short distance $0\nu\beta\beta$ operator should be absent, otherwise the tree-level contributions from the dim-11 $0\nu\beta\beta$ operators would be some minor corrections. In the present work, the absence of the lower order contribution is due to the absence of fields which generate $0\nu\beta\beta$ at lower order. That is to say the messenger fields of the dim-11 $0\nu\beta\beta$ operators are required to not mediate the lower dim-9 $0\nu\beta\beta$ operators. Notice that the lower order contribution could also be forbidden by introducing additional discrete or gauge symmetries, however, the resulting models would be much more complicated. It is well-known that the Majorana neutrino mass can be generated by the interactions of $0\nu\beta\beta$ decay~\cite{Schechter:1981bd}, the exchange of light Majorana neutrino can further contribute to $0\nu\beta\beta$ decay. The Majorana neutrino masses can be effectively described by the Weinberg operator. If the light Majorana neutrino masses are generated at tree level or one-loop level, the mediators should be very heavy or their couplings with SM fields should be very tiny in order to naturally accommodate the tiny neutrino masses. As a consequence, the contribution of the dim-11 effective operators to $0\nu\beta\beta$ could be highly suppressed. Hence we require that the tree level UV completion of the dim-11 $0\nu\beta\beta$ operators should not contain messengers which can mediate Weinberg operator through tree and one-loop Feynman diagram. From Eq.~\eqref{eq:0nbb-d11-opers}, we see that the dim-11 $0\nu\beta\beta$ operators involve either two Higgs fields $HH$ or their complex conjugate $H^{*}H^{*}$. If there is a 3-point vertex coupling two Higgs bosons with a new scalar in the $0\nu\beta\beta$ decay diagrams of figure~\ref{fig:genuine-diagram}, the quantum number of the scalar would be fixed to be $(\bm{1}, \bm{3}, -1, S)$ or its complex conjugate which is exactly the mediator of type-\uppercase\expandafter{\romannumeral2} seesaw mechanism. Therefore we have to exclude the diagrams which contain a 3-point vertex involving two external scalar lines. There are seven $0\nu\beta\beta$ diagrams in which the two external Higgs fields are attached to a single 3-point vertex, as shown in figure~\ref{fig:type-II-seesaw-nongenuine}. Thus there are totally 28 diagrams shown in figure~\ref{fig:genuine-diagrams} from which we can systematically find out the possible tree-level decompositions of all the dim-11 $0\nu\beta\beta$ operators $\mathcal{O}_{1a, 1b}$, $\mathcal{O}_{2a, 2b}$, $\mathcal{O}_{3a, 3b}$, $\mathcal{O}_4$, $\mathcal{O}_{5a, 5b, 5c, 5d}$, $\mathcal{O}_{6a, 6b}$ and $\mathcal{O}_{7a, 7b, 7c, 7d, 7e, 7f}$ in Eq.~\eqref{eq:0nbb-d11-opers}.

For each diagram in figure~\ref{fig:genuine-diagrams}, one has multiple choices for assigning the fields of any given dim-11 $0\nu\beta\beta$ operator to the external legs. Once a particular assignment is chosen, the SM
gauge charges of the internal fields can be fixed from gauge invariance of every interaction vertex. Then a diagram would be promoted to a well defined physical model which we will call a model-diagram or simply model. All the possible model-diagrams are listed in the attachment~\cite{Li:2026dim11NDBDsup}. In some specific models, some of the mediators carry the same SM gauge charges with identical Lorentz nature so that they can be identified as the same field.  After merging all the same fields, we get 487 model-diagrams with 3 internal fields, 2065 model-diagrams with 4 internal fields and 2728 model-diagrams with 5 internal fields. The numbers of model-diagrams for different $0\nu\beta\beta$ operators and topologies are shown in table~\ref{tab:num-of-model-diagram}. All these model-diagrams only involve 61 new field beyond SM which are labeled by some number from 1 to 61, as shown in table~\ref{tab:allnf}. We see that the generally these $0\nu\beta\beta$ decay models require fractionally charged fermions and exotic bosons which can be dileptons, diquarks or leptoquarks.

\begin{table}[t!]
\centering
\begin{tabular}{|c|c|c|c|c|c|c|c|}\hline\hline
\diagbox{Operator}{Topology}&T1&T2&T3&T4&T5&T6&T7\\ \hline
$\mathcal{O}_{1a}$, $\mathcal{O}_{1b}$&33& 27& 8& 335& 339& 26& 863\\\hline
$\mathcal{O}_{2a}$, $\mathcal{O}_{2b}$&49& 30& 9& 253& 204& 34& 354\\\hline
$\mathcal{O}_{3a}$, $\mathcal{O}_{3b}$&17& 8& 3& 119& 100& 15& 184\\\hline
$\mathcal{O}_4$&11& 0& 0& 83& 37& 2& 92\\\hline
$\mathcal{O}_{5a}$, $\mathcal{O}_{5b}$, $\mathcal{O}_{5c}$, $\mathcal{O}_{5d}$&9& 10& 5& 4& 45& 8& 60\\\hline
$\mathcal{O}_{6a}$, $\mathcal{O}_{6b}$&34& 27& 8& 116& 147& 0& 463\\\hline
$\mathcal{O}_{7a}$, $\mathcal{O}_{7b}$, $\mathcal{O}_{7c}$&\multirow{2}{*}{46}&\multirow{2}{*}{43}&\multirow{2}{*}{13}&\multirow{2}{*}{174}&\multirow{2}{*}{201}&\multirow{2}{*}{23}&\multirow{2}{*}{609}\\[-2mm]
$\mathcal{O}_{7d}$, $\mathcal{O}_{7e}$, $\mathcal{O}_{7f}$&&&&&&&\\
\hline\hline
\end{tabular}
\caption{\label{tab:num-of-model-diagram}The numbers of model-diagram for different $0\nu\beta\beta$ operators and topologies. Note that the topology T8 is not considered here because it always has the vertex with two Higgs bosons and a new scalar and consequently the contribution of type-II seesaw to neutrino masses can not be forbidden. }
\end{table}

It is notable that different model-diagrams would have the same set of messenger fields. Thus we combine the model-diagrams of the same mediator fields into a class named as model-fields, we find that there are 134 model-fields with 3 internal fields, 812 model-fields with 4 internal fields and 2075 model-fields with 5 internal fields. If any of the mediators is removed from a model, the dim-11 $0\nu\beta\beta$ operators can not be generated anymore, such a model would be regarded as ``minimal''. Obviously all these model-fields with 3 mediators are ``minimal'' models. All the ``minimal'' model-fields together with the corresponding model diagrams are listed in the attachment~\cite{Li:2026dim11NDBDsup}, including additional 275 model-fields with 4 mediators and 301 model-fields with 5 mediators.

\begin{table}[t!]
\centering
\renewcommand{\arraystretch}{0.7}
\scalebox{0.67}{
\begin{tabular}{|lccccccccc|}	
\hline\hline
		\rowcolor[HTML]{C0C0C0}{\color[HTML]{000000}}No.&1&\cellcolor[HTML]{CD9934}2&3&4&5&\cellcolor[HTML]{CD9934}6&\cellcolor[HTML]{CD9934}7&8&9\\
Irrep&$(\bm{1}, \bm{1}, -2, S)$&\cellcolor[HTML]{FFCB2F}$(\bm{1}, \bm{1}, -1, F)$ & $(\bm{1}, \bm{1}, -1,S)$ & $(\bm{1}, \bm{2}, -\frac{3}{2}, F)$ & $(\bm{1}, \bm{2}, -\frac{3}{2}, S)$&\cellcolor[HTML]{FFCB2F}$(\bm{1}, \bm{2}, -\frac{1}{2},F)$&\cellcolor[HTML]{FFCB2F}$(\bm{1}, \bm{2}, -\frac{1}{2},S)$&$(\bm{1}, \bm{3}, -2, S)$ & $(\bm{1}, \bm{3},-1,F)$\\[1mm]
\hline
&&&&&&&&&\\[-0.3cm]
\hline
		\rowcolor[HTML]{C0C0C0}{\color[HTML]{000000}}No.&10&11&\cellcolor[HTML]{CD9934}12&13&\cellcolor[HTML]{CD9934}14&15&16&17&18\\
Irrep&$(\bm{3}, \bm{1}, -\frac{4}{3}, F)$ & $(\bm{3}, \bm{1}, -\frac{4}{3}, S)$ &\cellcolor[HTML]{FFCB2F} $(\bm{3}, \bm{1}, -\frac{1}{3},F)$ & $(\bm{3}, \bm{1}, -\frac{1}{3}, S)$ &\cellcolor[HTML]{FFCB2F} $(\bm{3}, \bm{1}, \frac{2}{3}, F)$ & $(\bm{3}, \bm{1}, \frac{2}{3},S)$ & $(\bm{3}, \bm{1}, \frac{5}{3}, F)$ & $(\bm{3}, \bm{2}, -\frac{11}{6}, F)$ & $(\bm{3}, \bm{2}, -\frac{11}{6},S)$\\[1mm] \hline &&&&&&&&&\\[-0.3cm]
\hline
		\rowcolor[HTML]{C0C0C0}{\color[HTML]{000000}}No.&19&20&\cellcolor[HTML]{CD9934}21&22&23&24&25&26&27\\
Irrep & $(\bm{3}, \bm{2}, -\frac{5}{6}, F)$ & $(\bm{3}, \bm{2}, -\frac{5}{6}, S)$ & \cellcolor[HTML]{FFCB2F} $(\bm{3}, \bm{2}, \frac{1}{6}, F)$ & $(\bm{3}, \bm{1}, \frac{1}{6}, S)$ & $(\bm{3}, \bm{2}, \frac{7}{6}, F)$ & $(\bm{3}, \bm{2},\frac{7}{6}, S)$ & $(\bm{3}, \bm{2}, \frac{13}{6},F)$ & $(\bm{3}, \bm{3}, -\frac{7}{3}, S)$ & $(\bm{3}, \bm{3}, -\frac{4}{3}, F)$\\[1mm] \hline
&&&&&&&&&\\[-0.3cm]
\hline
		\rowcolor[HTML]{C0C0C0}{\color[HTML]{000000}}No.&28&29&30&31&32&33&34&35&36\\
Irrep&$(\bm{3}, \bm{3}, -\frac{4}{3}, S)$ & $(\bm{3}, \bm{3}, -\frac{1}{3}, F)$ & $(\bm{3}, \bm{3}, -\frac{1}{3}, S)$ & $(\bm{3}, \bm{3}, \frac{2}{3}, F)$ & $(\bm{3}, \bm{3}, \frac{2}{3}, S)$ & $(\bm{3}, \bm{3}, \frac{5}{3},F)$ & $(\bm{3}, \bm{3}, \frac{5}{3}, S)$ & $(\bm{6}, \bm{1}, -\frac{2}{3}, S)$ & $(\bm{6}, \bm{1}, \frac{1}{3}, F)$\\[1mm] \hline &&&&&&&&&\\[-0.3cm]
\hline
		
\rowcolor[HTML]{C0C0C0}{\color[HTML]{000000}}No.&37&38&39&40&41&42&43&44&45\\
Irrep& $(\bm{6}, \bm{1}, \frac{4}{3}, S)$ & $(\bm{6}, \bm{2}, -\frac{7}{6}, S)$ & $(\bm{6}, \bm{2}, -\frac{1}{6}, F)$ & $(\bm{6}, \bm{2}, -\frac{1}{6}, S)$ & $(\bm{6}, \bm{2}, \frac{5}{6}, F)$ & $(\bm{6}, \bm{2}, \frac{5}{6}, S)$ & $(\bm{6}, \bm{2}, \frac{11}{6}, S)$ & $(\bm{6}, \bm{3}, -\frac{5}{3}, S)$ & $(\bm{6}, \bm{2}, -\frac{2}{3}, F)$\\[1mm] \hline &&&&&&&&&\\[-0.3cm]
\hline
		
\rowcolor[HTML]{C0C0C0}{\color[HTML]{000000}}No.&46&47&48&49&50&51&52&53&54\\
Irrep & $(\bm{6}, \bm{3}, -\frac{2}{3}, S)$ & $(\bm{6}, \bm{3}, \frac{1}{3}, F)$ & $(\bm{6}, \bm{3}, \frac{1}{3}, S)$ & $(\bm{6}, \bm{3}, \frac{4}{3}, F)$ & $(\bm{6}, \bm{3}, \frac{4}{3}, S)$ & $(\bm{6}, \bm{3}, \frac{7}{3}, S)$ & $(\bm{8}, \bm{1}, -1, S)$ & $(\bm{8},\bm{1}, 0, F)$ & $(\bm{8}, \bm{2}, -\frac{3}{2}, S)$\\[1mm] \hline &&&&&&&&&\\[-0.3cm] \hline
		
\rowcolor[HTML]{C0C0C0}{\color[HTML]{000000}}No.&55&56&57&58&59&60&61&&\\
Irrep & $(\bm{8}, \bm{2}, -\frac{1}{2}, F)$ & $(\bm{8}, \bm{2}, -\frac{1}{2}, S)$ & $(\bm{8}, \bm{3}, -2, S)$ & $(\bm{8}, \bm{3}, -1, F)$ & $(\bm{8}, \bm{3}, -1, S)$ & $(\bm{8}, \bm{3}, 0, F)$ & $(\bm{8}, \bm{3}, 0, S)$ & &\\[1mm] \hline\hline
\end{tabular}}
\caption{\label{tab:allnf} The 61 new fields involved in the UV completions of the dime-11 $0\nu\beta\beta$ decay operators. We give the transformation of these new fields under the SM gauge group $SU(3)_C\times SU(2)_L\times U(1)_Y$, and the last symbol ``$S$'' and ``$F$'' refers to scalar and fermion respectively. The hypercharge $Y$ of a field is related to its electric charge $Q$ via the Gell-Mann-Nishijima formula $Q=T_3+Y$, where $T_3$ is the third component of the weak isospin. The fields in the yellow shaded cell transform in the same way as certain SM particles under the SM gauge group. Notice that all the new fermions are assumed to be vector-like fermions for anomaly cancellation while all SM fermions are chiral fermions. }
\end{table}

\section{\label{sec:representative-model}A representative $0\nu\beta\beta$ model with colorless mediator }

Almost all dim-11 $0\nu\beta\beta$ decompositions involve colored mediators. Only two realizations consist solely of three $SU(3)_C$-singlet fields and do not require any auxiliary symmetry to suppress lower-order contributions. One of these models is presented below, while the second is described in appendix~\ref{app:2nd-colorless-model}.

In this model, one needs to introduce the following three new scalar fields to the SM particle content,
\begin{eqnarray}
\label{eq:scalars-model1}\phi\equiv\begin{pmatrix}
\phi^+\\\phi^0
\end{pmatrix}\sim(\bm{1}, \bm{2}, 1/2, S)\,,~~~\zeta\equiv\zeta^{++}\sim(\bm{1}, \bm{1}, 2, S)\,,~~~\eta\equiv\begin{pmatrix}
\eta^{++}\\\eta^+
\end{pmatrix}\sim(\bm{1}, \bm{2}, 3/2, S)\,,
\end{eqnarray}
where $\phi$ is a diquark doublet, $\zeta$ is a dilepton singlet, and the scalar doublet $\eta$ carrying exotic electric charges. We see that the above fields $\zeta$, $\eta$ and $\phi$ are the complex conjugates of the 1st, 5th and 7th fields in table~\ref{tab:allnf}.
Notice that the scalar doublet $\phi$ carries the same SM gauge charge as the SM Higgs boson $H$, consequently it is the second Higgs doublet. In the unitary gauge, the SM Higgs field can be parameterized as
\begin{equation}
\label{eq:Higgs-VEV} H(x)=\frac{1}{\sqrt{2}}\begin{pmatrix}
0\\ v+h(x)
\end{pmatrix}\,,
\end{equation}
where $v\approx246$ GeV is the vacuum expectation value (VEV) of the neutral component of the Higgs doublet $H$~\cite{ParticleDataGroup:2024cfk}. Using the freedom of field redefinition, one can  always choose a basis in which the VEV of the second Higgs field $\phi^0$ is vanishing~\cite{Branco:1999fs}. We can straightforwardly read off the gauge invariant Yukawa interactions as follows,
\begin{eqnarray}
\label{eq:Lagrangian-model1}-\mathcal{L}_Y&=&-\mathcal{L}^{SM}_Y+h_{\alpha\beta}\,\overline{u_{R\alpha}}\phi^Ti\sigma_2Q_{L\beta}+h'_{\alpha\beta}\,\overline{Q_{L\alpha}}\phi d_{R\beta}+y_{\alpha\beta}\,\overline{e_{R\alpha}^C}\zeta e_{R\beta}+h''_{\alpha\beta}\,\overline{\ell_{L\alpha}}\phi e_{R\beta}+\text{h.c.}\,,~~~~
\end{eqnarray}
where $\mathcal{L}^{SM}_Y$ denotes the SM Yukawa couplings, and $\alpha,\beta=1,2,3$ are generation indices, $\sigma_2=\begin{pmatrix}
0 ~& -i \\
i ~& 0\end{pmatrix}$ is the second Pauli matrix. The coupling $y_{\alpha\beta}=y_{\beta\alpha}$ is symmetric with respect to the flavor indices $\alpha$ and $\beta$. Notice that only the interactions of the first generation fermion are relevant to $0\nu\beta\beta$ decay with $\alpha=\beta=1$.

The introduction of the three complex scalar make the scalar potential $V$ quite complicated and there are a large number of terms compatible with SM gauge symmetry,
\begin{eqnarray}
V&=&-\mu^2H^{\dagger}H+m_\phi^2\phi^\dagger\phi+m_{\zeta}^2\zeta^{\dagger}\zeta+m_{\eta}^2\eta^\dagger\eta+\Big(m^2\phi^\dagger H+\text{h.c.}\Big)+\left(\mu_1\,H^\dagger\widetilde{\eta}\zeta+\mu_2\,\phi^{\dagger}\widetilde{\eta}\zeta+\text{h.c.}\right)\nonumber\\
&&+\lambda\left(H^{\dagger}H\right)^2+\lambda_1\left(\eta^\dagger\eta\right)^2+\lambda_2\left(\eta^\dagger\eta\right)\left(\zeta^{\dagger}\zeta\right)
+\lambda_3\left(\phi^\dagger\phi\right)^2+\lambda_4\left(\eta^{\dagger}\eta\right)\left(\phi^\dagger\phi\right)
+\lambda_5\left(\eta^\dagger\phi\right)\left(\phi^{\dagger}\eta\right)\nonumber\\
&&+\lambda_6\left(\eta^\dagger\eta\right)\left(H^{\dagger}H\right)+\lambda_7\left(\eta^\dagger H\right)\left(H^\dagger\eta\right)+\lambda_8\left(\zeta^{\dagger}\zeta\right)^2+\lambda_9\left(H^\dagger H\right)\zeta^{\dagger}\zeta+\lambda_{10}\left(\phi^\dagger \phi\right)\zeta^{\dagger}\zeta\nonumber\\
&&+\lambda_{11}\left(\phi^{\dagger}H\right)\left(H^\dagger\phi\right)+\lambda_{12}\left(\phi^{\dagger}\phi\right)
\left(H^{\dagger}H\right)+\Big[\lambda_{13}\left(H^\dagger\widetilde{\phi}\right)\left(\phi^\dagger\eta\right)
+\lambda_{14}\left(H^{\dagger}\widetilde{\phi}\right)\left(H^{\dagger}\eta\right)\nonumber\\
&&+\lambda_{15}\left(\phi^{\dagger}H\right)\left(\eta^{\dagger}\eta\right)+\lambda_{16}\left(\phi^{\dagger}H\right)^2
+\lambda_{17}\left(\eta^{\dagger}H\right)\left(\phi^\dagger\eta\right)+\lambda_{18}\left(\eta^\dagger\eta\right)\left(\phi^{\dagger} H\right)+\lambda_{19}\left(\phi^{\dagger}H\right)\zeta^{\dagger}\zeta\nonumber\\
\label{eq:scalar-poten-model1}&&+\lambda_{20}\left(\phi^{\dagger}\phi\right)\left(\phi^{\dagger}H\right)+\lambda_{21}
\left(\phi^{\dagger}H\right)\left(H^{\dagger}H\right)+\text{h.c.}\Big]\,,
\end{eqnarray}
where $\widetilde{\eta}=i\sigma_2\eta^{*}$ and $\widetilde{\phi}=i\sigma_2\phi^{*}$.
The parameter $m_H$, $m_\phi$, $m_\zeta$, $m_\eta$, $m$, $\mu_1$ and $\mu_2$ have dimension of mass, while the couplings $\lambda$ and $\lambda_i$ with $i=1,2,\ldots, 21$ are dimensionless. The parameters $m$, $\mu_1$, $\mu_2$ and $\lambda_j$ ($j=13, 14\ldots, 21$) are generally complex, while the other coupling constants are real. Moreover, the complex phases of $\mu_1$, $\lambda_{14}$ and $\lambda_{16}$ could be absorbed by redefinition of $\zeta$, $\phi$ and $\eta$ respectively. Note that $\phi$ can't be identified as the SM Higgs boson $H$ otherwise the interaction terms $\lambda_{13}\,(H^\dagger\widetilde{\phi})(\phi^\dagger\eta)$ and $\lambda_{14}\,(H^\dagger\widetilde{\phi})(H^\dagger\eta)$ would be absent and the $0\nu\beta\beta$ decay can not be mediated, as can be seen from figure~\ref{fig:model-su3sing-3s} and explained later. Furthermore, the minimization condition of the potential at the vacuum $\langle H\rangle=\left(0, v/\sqrt{2}\right)^T$, $\langle\phi\rangle=\left(0, 0\right)^T$ leads to
\begin{eqnarray}
\mu^2&=&\lambda v^2\,,~~~m^2=-\frac{1}{2}\lambda_{21}v^2\,.\label{eq:mf3i5-subeq}
\end{eqnarray}

\subsection{The mass eigenstates and mass spectrum of scalar fields }

From Eq.~\eqref{eq:scalars-model1}, we see that the scalar spectrum of the model contains two doubly-charged scalars $\eta^{++}$, $\zeta^{++}$ and their complex conjugates, and two singly-charged scalars $\phi^{+}$, $\eta^{+}$ and their conjugates. Inserting the SM Higgs field in Eq.~\eqref{eq:Higgs-VEV} into the scalar potential $V$ of Eq.~\eqref{eq:scalar-poten-model1}, we can read off the quadratic terms for the charged scalars as follows,
\begin{eqnarray}
\mathcal{L}^{\text{CS}}_{mass}&=&-\left(m_\zeta^2+\frac{\lambda_9v^2}{2}\right)\zeta^{--}\zeta^{++}-\left(m_\eta^2+\frac{\lambda_6v^2}{2}\right)\eta^{--}\eta^{++}
+\frac{\mu_1v}{\sqrt{2}}\eta^{--}\zeta^{++}+\frac{\mu_1^*v}{\sqrt{2}}\eta^{++}\zeta^{--}\nonumber\\
&&\hskip-0.2in -\left(m_\phi^2+\frac{\lambda_{12}v^2}{2}\right)\phi^-\phi^{+}-\left(m_\eta^2+\frac{\lambda_6v^2}{2}-\frac{\lambda_7v^2}{2}\right)\eta^-\eta^+
+\frac{\lambda_{14}v^2}{2}\phi^{-}\eta^{+}+\frac{\lambda_{14}^*v^2}{2}\phi^+\eta^{-}\,,\label{eq:mf3i5-mix-char}
\end{eqnarray}
which is responsible for the masses of charged scalars with $\zeta^{--}=(\zeta^{++})^{*}$, $\eta^{--}=(\eta^{++})^{*}$, $\phi^{-}=(\phi^{+})^{*}$ and $\eta^{-}=(\eta^{+})^{*}$. The parameters $\mu_1$ and $\lambda_{14}$ can be taken real by redefining the fields $\zeta$ and $\eta$. The mass terms for $\zeta^{++}$ and $\eta^{++}$ can be written in matrix form as
\begin{eqnarray}
-\left(\zeta^{--},\,\eta^{--}\right)\begin{pmatrix}
m_{\zeta}^2+\frac{\lambda_9v^2}{2}~&~-\frac{\mu_1 v}{\sqrt{2}}\\
-\frac{\mu_1 v}{\sqrt{2}}~&~m_{\eta}^2+\frac{\lambda_6v^2}{2}
\end{pmatrix}\begin{pmatrix}
\zeta^{++}\\\eta^{++}
\end{pmatrix}\,,
\end{eqnarray}
where the off-diagonal entry $-\frac{\mu_1 v}{\sqrt{2}}$ mixes $\zeta^{++}$ and $\eta^{++}$. The mass eigenstates of doubly-charged scalars, which are denoted as $\varsigma^{++}_1$ and $\varsigma^{++}_2$, are linear combinations of $\zeta^{++}$ and $\eta^{++}$,
\begin{eqnarray}
\begin{pmatrix}
\varsigma_1^{++}\\\varsigma_2^{++}
\end{pmatrix}
&=&\begin{pmatrix}
\cos\theta_\varsigma~&~\sin\theta_\varsigma\\
-\sin\theta_\varsigma~&~\cos\theta_\varsigma
\end{pmatrix}\begin{pmatrix}
\zeta^{++}\\\eta^{++}
\end{pmatrix}
\label{eq:mix-eta12++}\,,
\end{eqnarray}
where the rotation angle $\theta_\varsigma$ is
\begin{eqnarray}
\tan2\theta_\varsigma&=&\frac{-\sqrt{2}\mu_1 v}{\hat{m}_{\zeta}^2-\hat{m}_{\eta}^2},~ \sin2\theta_\varsigma=\frac{-\sqrt{2}\mu_1 v}{\sqrt{\left(\hat{m}_{\zeta}^2-\hat{m}_{\eta}^2\right)^2+2\mu_1 ^2v^2}},~ \cos2\theta_\varsigma=\frac{\hat{m}_{\zeta}^2-\hat{m}_{\eta}^2}{\sqrt{\left(\hat{m}_{\zeta}^2-\hat{m}_{\eta}^2\right)^2+2\mu_1 ^2v^2}}\,.
 \nonumber\\
\end{eqnarray}
The mass eigenvalues of determined to be
\begin{eqnarray}
\nonumber m_{\varsigma_1}^2&=&\frac{1}{2}\Big[\hat{m}_{\zeta}^2+\hat{m}_{\eta}^2+\sqrt{\left(\hat{m}_{\zeta}^2-\hat{m}_{\eta}^2\right)^2+2\mu_1 ^2v^2}\Big],\\
m_{\varsigma_2}^2&=&\frac{1}{2}\Big[\hat{m}_{\zeta}^2+\hat{m}_{\eta}^2-\sqrt{\left(\hat{m}_{\zeta}^2-\hat{m}_{\eta}^2\right)^2+2\mu_1 ^2v^2}\Big]\,,
\end{eqnarray}
with
\begin{equation}
\hat{m}_\zeta^2\equiv m_{\zeta}^2+\frac{\lambda_9v^2}{2},\qquad \hat{m}_\eta^2\equiv m_{\eta}^2+\frac{\lambda_6v^2}{2}\,.
\end{equation}
One sees that the mass splitting between $\varsigma_1$ and $\varsigma_2$ fulfills
$m_{\varsigma_1}^2-m_{\varsigma_2}^2=\sqrt{\left(\hat{m}_{\zeta}^2-\hat{m}_{\eta}^2\right)^2+2\mu_1 ^2v^2}$ which depends on the parameters $\lambda_6$, $\lambda_9$ and $\mu_1$ as well as the Higgs VEV $v$. This mass splitting is crucially relevant to the light neutrino mass generation, as shown in the following subsection. Analogously the mass eigenstates of singly-charged scalars, denoted as $\kappa^+_1$ and $\kappa^+_2$, are linear combinations of $\eta^{+}$ and $\phi^{+}$,
\begin{eqnarray}
\begin{pmatrix}
\eta^+\\\phi^+
\end{pmatrix}=\begin{pmatrix}
\cos\theta_\kappa~&~-\sin\theta_\kappa\\
\sin\theta_\kappa~&~\cos\theta_\kappa
\end{pmatrix}\begin{pmatrix}
\kappa^+_1\\\kappa^+_2
\end{pmatrix}\label{eq:mix-eta2ph2p}\,,
\end{eqnarray}
with the rotation angle $\theta_\kappa$
\begin{eqnarray}
\tan2\theta_\kappa&=&\frac{-\lambda_{14} v^2}{\widetilde{m}_{\eta}^2-\widetilde{m}_{\phi}^2},~ \sin2\theta_\kappa=\frac{-\lambda_{14} v^2}{\sqrt{\left(\widetilde{m}_{\eta}^2-\widetilde{m}_{\phi}^2\right)^2+\lambda_{14}^2v^4}},~ \cos2\theta_\kappa=\frac{\widetilde{m}_{\eta}^2-\widetilde{m}_{\phi}^2}{\sqrt{\left(\widetilde{m}_{\eta}^2-\widetilde{m}_{\phi}^2\right)^2+\lambda_{14}^2v^4}}\,.
 \nonumber\\
\end{eqnarray}
The mass eigenvalues of $\kappa^+_1$ and $\kappa^+_2$ are
\begin{eqnarray}
\nonumber m_{\kappa_1}^2&=&\frac{1}{2}\Big[\widetilde{m}_{\eta}^2+\widetilde{m}_{\phi}^2+\sqrt{\left(\widetilde{m}_{\eta}^2-\widetilde{m}_{\phi}^2\right)^2+\lambda_{14}^2v^4}\Big],\\ m_{\kappa_2}^2&=&\frac{1}{2}\Big[\widetilde{m}_{\eta}^2+\widetilde{m}_{\phi}^2-\sqrt{\left(\widetilde{m}_{\eta}^2-\widetilde{m}_{\phi}^2\right)^2+\lambda_{14}^2v^4}\Big]\,,
\end{eqnarray}
with
\begin{equation}
\widetilde{m}_\eta^2\equiv m_\eta^2+\frac{\lambda_6v^2}{2}+\frac{\lambda_7v^2}{2},\qquad \widetilde{m}_\phi^2\equiv m_\phi^2+\frac{\lambda_{12}v^2}{2}\,.
\end{equation}
The mass splitting between $\kappa^+_1$ and $\kappa^+_2$ depends on the couplings $\lambda_6$, $\lambda_7$, $\lambda_{12}$ and $\lambda_{14}$, and it plays a critical role in neutrino mass generation. Now we proceed to consider the neutral scalars. The field $\phi^0$ which is the second component of the doublet $\phi$, is a complex scalar, consequently it can be written as $\phi^0=\frac{1}{\sqrt{2}}(\phi_R+i\phi_I)$. Hence this model contains three chargeless scalars $h$, $\phi_R$ and $\phi_I$. The quadratic mass terms of these three neutral scalars can be read out from the scalar potential as,
\begin{eqnarray}
\mathcal{L}^{\text{NS}}_{mass}&=&-\lambda v^2h^2-\frac{1}{2}\left[m_\phi^2+\frac{\lambda_{11}v^2}{2}+\frac{\lambda_{12}v^2}{2}+\text{Re}\left(\lambda_{16}\right)v^2\right]\phi_R^2\nonumber\\
&&-\frac{1}{2}\left[m_\phi^2+\frac{\lambda_{11}v^2}{2}+\frac{\lambda_{12}v^2}{2}-\text{Re}\left(\lambda_{16}\right)v^2\right]\phi_I^2
-\text{Re}\left(\lambda_{21}\right)v^2h\phi_R\nonumber\\
&&-\text{Im}\left(\lambda_{21}\right)v^2h\phi_I-\text{Im}\left(\lambda_{16}\right)v^2\phi_R\phi_I\nonumber\\
&=&-\frac{1}{2}\left(h~\phi_R~\phi_I\right)\mathcal{M}^2\begin{pmatrix}
h\\\phi_R\\\phi_I
\end{pmatrix}\,,
\end{eqnarray}
where
\begin{eqnarray}
\mathcal{M}^2&\equiv&\begin{pmatrix}
2\lambda v^2~&~\text{Re}\left(\lambda_{21}\right)v^2~&~\text{Im}\left(\lambda_{21}\right)v^2\\
\text{Re}\left(\lambda_{21}\right)v^2~&~\widetilde{m}_\phi^2+\frac{\lambda_{11}v^2}{2}+\text{Re}\left(\lambda_{16}\right)v^2~&~\text{Im}\left(\lambda_{16}\right)v^2\\
\text{Im}\left(\lambda_{21}\right)v^2~&~\text{Im}\left(\lambda_{16}\right)v^2~&~\widetilde{m}_\phi^2+\frac{\lambda_{11}v^2}{2}-\text{Re}\left(\lambda_{16}\right)v^2
\end{pmatrix}\,.
\end{eqnarray}
The mass eigenstates of neutral scalars and their physical masses can be obtained by diagonalizing the symmetric matrix $\mathcal{M}^2$. The corresponding expressions are too lengthy to provide some useful insight, consequently we don't present them here. By exploiting the rephasing freedom of $\phi$, one may set either $\lambda_{16}$ or $\lambda_{21}$ to be real, while the other is generally complex. For instance one can choose $\lambda_{16}$ to be real and $\lambda_{21}$ complex. In the limit both $\lambda_{16}$ or $\lambda_{21}$ are real, the mass matrix $\mathcal{M}^2$ would be block diagonal,
\begin{eqnarray}
\mathcal{M}^2&\equiv&\begin{pmatrix}
2\lambda v^2~&~\lambda_{21}v^2~&~0\\
\lambda_{21}v^2~&~\widetilde{m}_\phi^2+\frac{\lambda_{11}v^2}{2}+\lambda_{16}v^2~&~0\\
0~&~0~&~\widetilde{m}_\phi^2+\frac{\lambda_{11}v^2}{2}-\lambda_{16}v^2
\end{pmatrix}\,.
\end{eqnarray}
We see that $h$ and $\phi_R$ would be mixed with each other, and the linear combination of them gives rise to the mass eigenstates:
\begin{eqnarray}
\begin{pmatrix}
h\\ \phi_R
\end{pmatrix}&=&\begin{pmatrix}
\cos\theta~&~-\sin\theta\\
\sin\theta~&~\cos\theta
\end{pmatrix}\begin{pmatrix}
S_1 \\ S_2
\end{pmatrix}
\label{eq:mix-h-phiR}\,,
\end{eqnarray}
where the rotation angle $\theta$ is
\begin{eqnarray}
\tan2\theta&=&\frac{2\lambda_{21}v^2}{m_h^2-m_R^2},\quad \sin2\theta=\frac{2\lambda_{21}v^2}{\sqrt{\left(m_h^2-m_R^2\right)^2+4\lambda_{21}^2v^4}},\quad \cos2\theta=\frac{m_h^2-m_R^2}{\sqrt{\left(m_h^2-m_R^2\right)^2+4\lambda_{21}^2v^4}},
\nonumber\\
\end{eqnarray}
The mass eigenvalues of $m_{S_1}$ and $m_{S_2}$ are given by
\begin{eqnarray}	
\nonumber m_{S_1}^2&=&\frac{1}{2}\Big[m_h^2+m_R^2+\sqrt{\left(m_h^2-m_R^2\right)^2+4\lambda_{21}^2v^4}\Big],\\ m_{S_2}^2&=&\frac{1}{2}\Big[m_h^2+m_R^2-\sqrt{\left(m_h^2-m_R^2\right)^2+4\lambda_{21}^2v^4}\Big]\,,
\end{eqnarray}
with
\begin{equation}
m_h^2\equiv 2\lambda v^2,\qquad m_R^2\equiv\widetilde{m}_\phi^2+\frac{\lambda_{11}v^2}{2}+\lambda_{16}v^2\,.
\end{equation}
After electroweak symmetry breaking, and expressing all fields in the mass-eigenstate basis, we obtain the following interactions relevant to neutrinoless double beta decay and light neutrino masses,
\begin{eqnarray}
-\mathcal{L}&=&\sum_{i=1,2}\left[f''_{i\alpha\beta}\kappa^{+}_{i}\overline{\nu_a}P_Re_\beta+y'_{i\alpha\beta}\varsigma_i^{++}\overline{e_\alpha^C}P_Re_\beta+
\overline{u_{\alpha}}\left(f_{i\alpha\beta}P_L+f'_{i\alpha\beta}P_R\right)d_{\beta}\kappa_i^+\right]\nonumber\\
&&-\text{i}\sum_{i,j=1,2}\lambda_{ij}\,W^{\mu+}\left(\kappa_i^{+}\partial_{\mu}\varsigma_j^{--}-\partial_{\mu}\kappa_i^{+}\varsigma_j^{--}\right)
+\sum_{i,j,k=1,2}\mu_{ijk}\kappa_i^{-}\kappa_{j}^{-}\varsigma_{k}^{++}+\text{h.c}\,,
\end{eqnarray}
where the couplings are
\begin{eqnarray}
\nonumber&& f''_{1\alpha\beta}=h''_{\alpha\beta}\sin\theta_\kappa\,,\quad f''_{2\alpha\beta}= h''_{\alpha\beta}\cos\theta_\kappa\,,\quad y'_{1\alpha\beta}= y_{\alpha\beta} \cos\theta_\varsigma\,,\quad y'_{2\alpha\beta}=-y_{\alpha\beta} \sin\theta_\varsigma\,,\nonumber\\
&&f_{1\alpha\beta}=h_{\alpha\beta}\sin\theta_\kappa\,,\quad f_{2\alpha\beta}= h_{\alpha\beta}\cos\theta_\kappa\,,\quad f'_{1\alpha\beta}=h'_{\alpha\beta}\sin\theta_\kappa\,,\quad f'_{2\alpha\beta}=h'_{\alpha\beta}\cos\theta_\kappa\,, \label{eq:f-h-fp-fpp-yp-def}
\end{eqnarray}
and
\begin{eqnarray}
\lambda_{11}=\frac{g\cos\theta_\kappa\sin\theta_\varsigma}{\sqrt{2}}\,,~
\lambda_{12}=\frac{g\cos\theta_\kappa\cos\theta_\varsigma}{\sqrt{2}}\,,~
\lambda_{21}=-\frac{g\sin\theta_\kappa\sin\theta_\varsigma}{\sqrt{2}}\,,
\lambda_{22}=-\frac{g\sin\theta_\kappa\cos\theta_\varsigma}{\sqrt{2}}\,,~~~\label{eq:lambda-ij}
\end{eqnarray}
and
\begin{eqnarray}
\mu_{111}&=&\sin\theta_\kappa\left(\mu_2\cos\theta_\kappa\cos\theta_\varsigma-\frac{\lambda_{13} v}{\sqrt{2}}\sin\theta_\kappa\sin\theta_\varsigma+\frac{\lambda_{17}v}{\sqrt{2}}\cos\theta_\kappa\sin\theta_\varsigma\right)\,,\nonumber\\
\mu_{112}&=&\sin\theta_\kappa\left(-\mu_2 \cos\theta_\kappa\sin\theta_\varsigma-\frac{\lambda_{13} v}{\sqrt{2}}\sin\theta_\kappa\cos\theta_\varsigma+\frac{\lambda_{17}v}{\sqrt{2}}\cos\theta_\kappa\cos\theta_\varsigma\right)\,,\nonumber\\
\mu_{121}=\mu_{211}&=&\frac{1}{2}\left(\mu_2 \cos2\theta_\kappa\cos\theta_\varsigma-\frac{\lambda_{13} v}{\sqrt{2}}\sin2\theta_\kappa\sin\theta_\varsigma+\frac{\lambda_{17}v}{\sqrt{2}}\cos2\theta_\kappa\sin\theta_\varsigma\right)\,,\nonumber\\
\mu_{122}=\mu_{212}&=&\frac{1}{2}\left(-\mu_2 \cos2\theta_\kappa\sin\theta_\varsigma-\frac{\lambda_{13} v}{\sqrt{2}}\sin2\theta_\kappa\cos\theta_\varsigma+\frac{\lambda_{17}v}{\sqrt{2}}\cos2\theta_\kappa\cos\theta_\varsigma\right)\,,\nonumber\\
\mu_{221}&=&\cos\theta_\kappa\left(-\mu_2 \sin\theta_\kappa\cos\theta_\varsigma-\frac{\lambda_{13}v}{\sqrt{2}}\cos\theta_\kappa\sin\theta_\varsigma-\frac{\lambda_{17}v}{\sqrt{2}}\sin\theta_\kappa\sin\theta_\varsigma\right)\,,\nonumber\\
\mu_{222}&=&\cos\theta_\kappa\left(\mu_2 \sin\theta_\kappa\sin\theta_\varsigma-\frac{\lambda_{13} v}{\sqrt{2}}\cos\theta_\kappa\cos\theta_\varsigma-\frac{\lambda_{17}v}{\sqrt{2}}\sin\theta_\kappa\cos\theta_\varsigma\right)\,.	\label{eq:mu-ijk}
\end{eqnarray}

\subsection{Half-time of $0\nu\beta\beta$ decays }

\begin{figure}[hptb]
\centering
\includegraphics[scale=0.62]{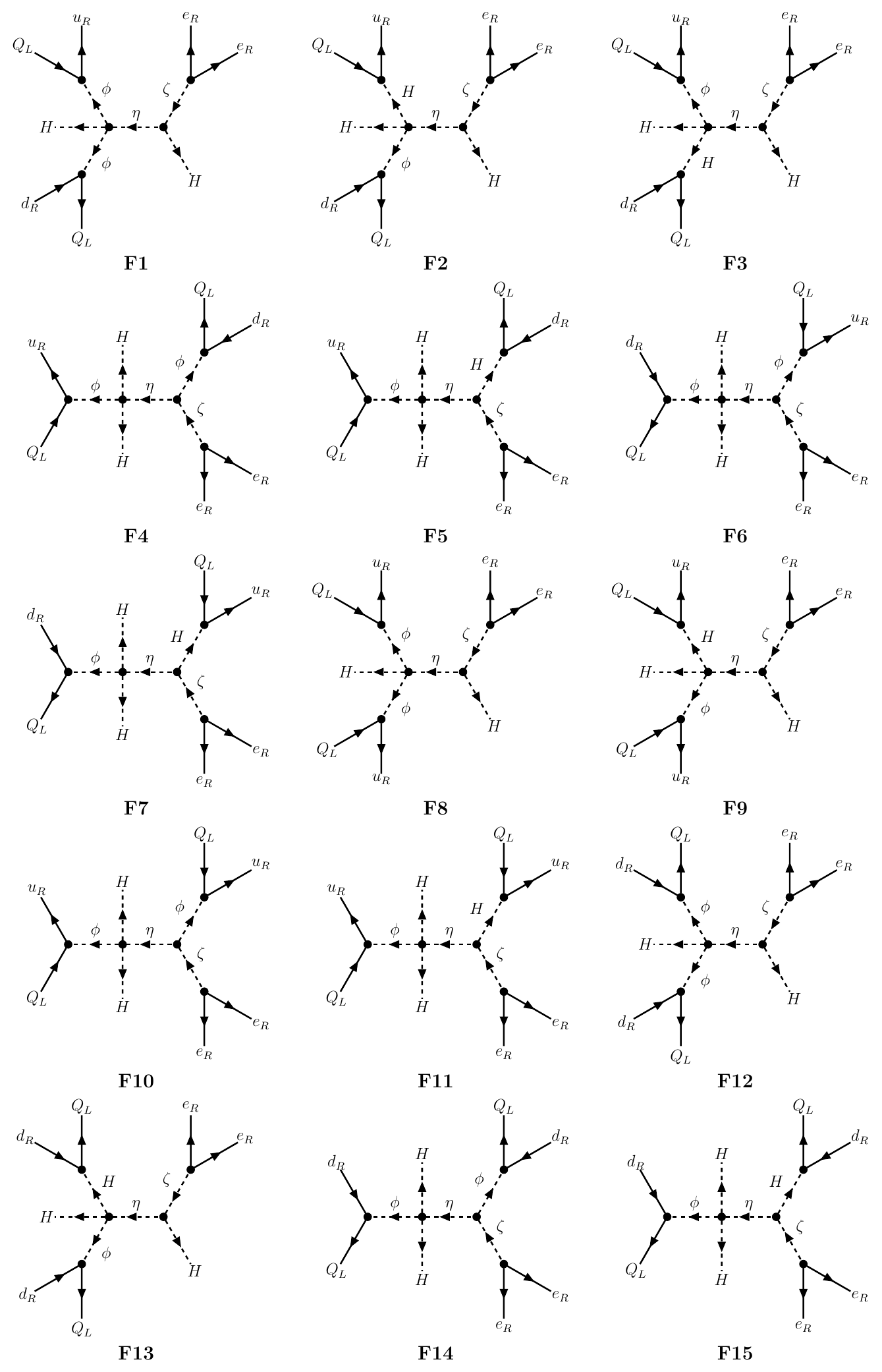}
\caption{\label{fig:model-su3sing-3s}The Feynman diagrams for the short-range $0\nu\beta\beta$ decay in the example model of section~\ref{sec:representative-model}. Notice that the panels of F1$\sim$F7 are UV completions of the $0\nu\beta\beta$ operators $\mathcal{O}_{1a}$ and $\mathcal{O}_{1b}$, F8$\sim$F11 are UV completions of $\mathcal{O}_{2a}$ and  $\mathcal{O}_{2b}$, F12$\sim$F15 are UV completions of the operators $\mathcal{O}_{3a}$ and $\mathcal{O}_{3b}$. }
\end{figure}

\begin{figure}[t!]
\centering
\includegraphics[scale=0.8]{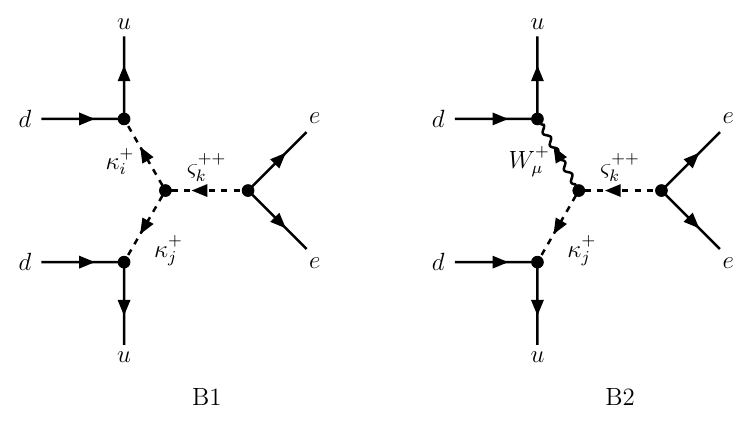}
\caption{\label{fig:model_SU3sing_3s_break}The Feynman diagrams of the $0\nu\beta\beta$ decay in the mass basis for the example model of section~\ref{sec:representative-model}, where the unitary gauge ia adopted. }
\end{figure}

\begin{figure}[t!]
\centering
\includegraphics[scale=0.65]{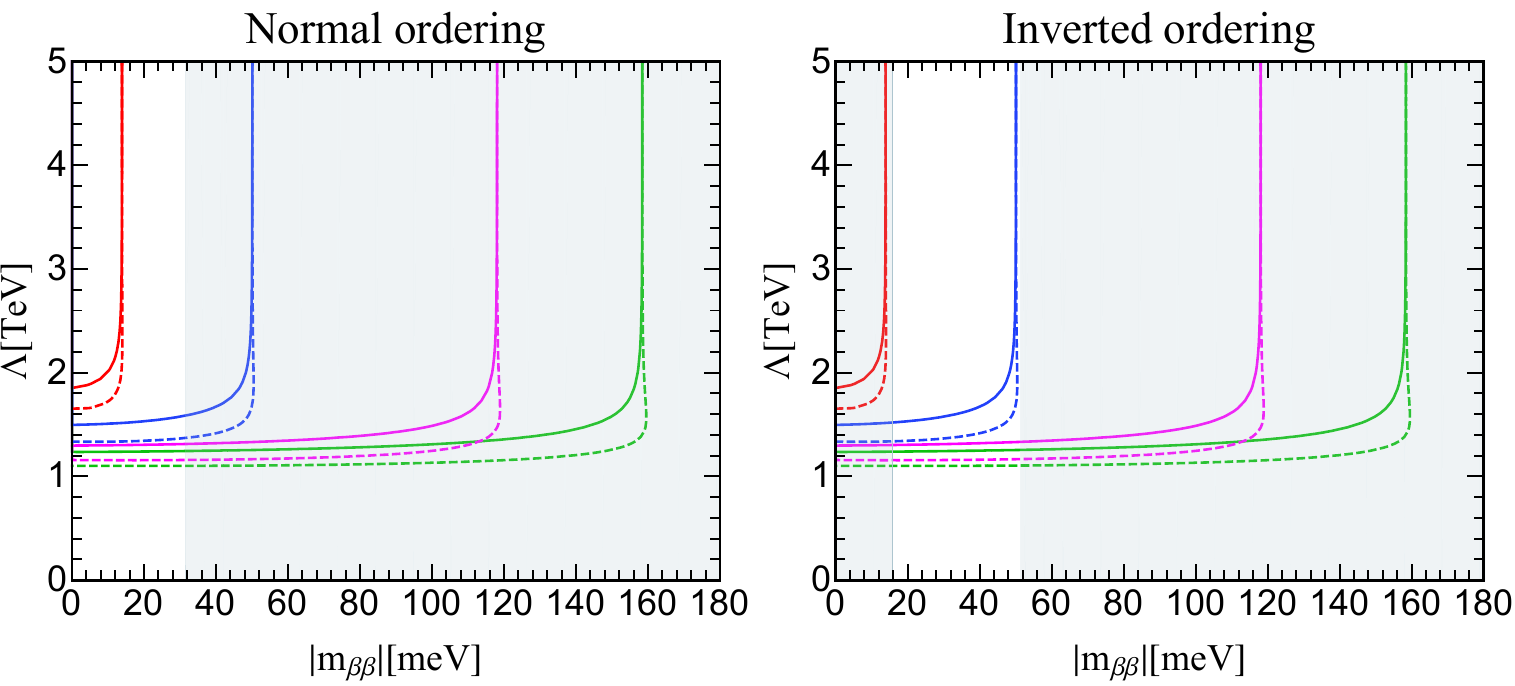}\\[-0.1in]
\qquad\qquad\qquad \resizebox{0.98\textwidth}{!}{\hspace{1cm}
\begin{tabular}{|ccccc|}\hline
$\texttt{T}_{1/2}^\texttt{Ge}$&$1.0\times10^{26}\texttt{ yr}$&$1.8\times10^{26}\texttt{ yr}$&$1.0\times10^{27}\texttt{ yr}$&$1.3\times10^{28}\texttt{ yr}$\\
$(\alpha,\beta,\gamma)=(0,0,\pi)$&\begin{tikzpicture}
\definecolor{greeen}{RGB}{12,194,14};
\draw[greeen,ultra thick] (-1.0,0) --(1.0,0);
\end{tikzpicture}&\begin{tikzpicture}
\definecolor{greeen}{RGB}{255,0,255};
\draw[greeen,ultra thick] (-1.0,0) --(1.0,0);
\end{tikzpicture}&\begin{tikzpicture}
\definecolor{greeen}{RGB}{34,65,250};
\draw[greeen,ultra thick] (-1.0,0) --(1.0,0);
\end{tikzpicture}&\begin{tikzpicture}
\definecolor{greeen}{RGB}{248,42,55};
\draw[greeen,ultra thick] (-1.0,0) --(1.0,0);
\end{tikzpicture}\\
$(\alpha,\beta,\gamma)=(0,\pi,0)$&\begin{tikzpicture}
\definecolor{greeen}{RGB}{12,194,14};
\draw[greeen,ultra thick,dashed] (-1.0,0) --(1.0,0);
\end{tikzpicture}&\begin{tikzpicture}
\definecolor{greeen}{RGB}{255,0,255};
\draw[greeen,ultra thick,dashed] (-1.0,0) --(1.0,0);
\end{tikzpicture}&\begin{tikzpicture}
\definecolor{greeen}{RGB}{34,65,250};
\draw[greeen,ultra thick,dashed] (-1.0,0) --(1.0,0);
\end{tikzpicture}&\begin{tikzpicture}
\definecolor{greeen}{RGB}{248,42,55};
\draw[greeen,ultra thick,dashed] (-1.0,0) --(1.0,0);
\end{tikzpicture}\\\hline
\end{tabular}\hspace{0.6cm}}
\caption{\label{fig:mf3i5-mbbGe-lambda}Contour plots of the
$0\nu\beta\beta$ decay half-life of $^{76}\text{Ge}$ in the plane $\Lambda$ versus $|m_{\beta\beta}|$ for NO and IO neutrino mass spectrums, where $|m_{\beta\beta}|$ is the effective Majorana neutrino mass and $\Lambda$ denotes the mass scale of new fields. The pink and red curves correspond to the current experimental limit $T_{1/2}(^{76}\text{Ge})=1.8\times10^{26}$ yr~\cite{GERDA:2020xhi} and the projected future sensitivity $T_{1/2}(^{76}\text{Ge})=1.3\times10^{28}$ yr~\cite{LEGEND:2021bnm}, respectively. The gray bands indicate the regions of $|m_{\beta\beta}|$ disfavored by the current neutrino oscillation data~\cite{Esteban:2024eli} and the cosmological bound $\sum_{i=1}^3 m_i<0.12$ eV~\cite{Planck:2018vyg}.}
\end{figure}
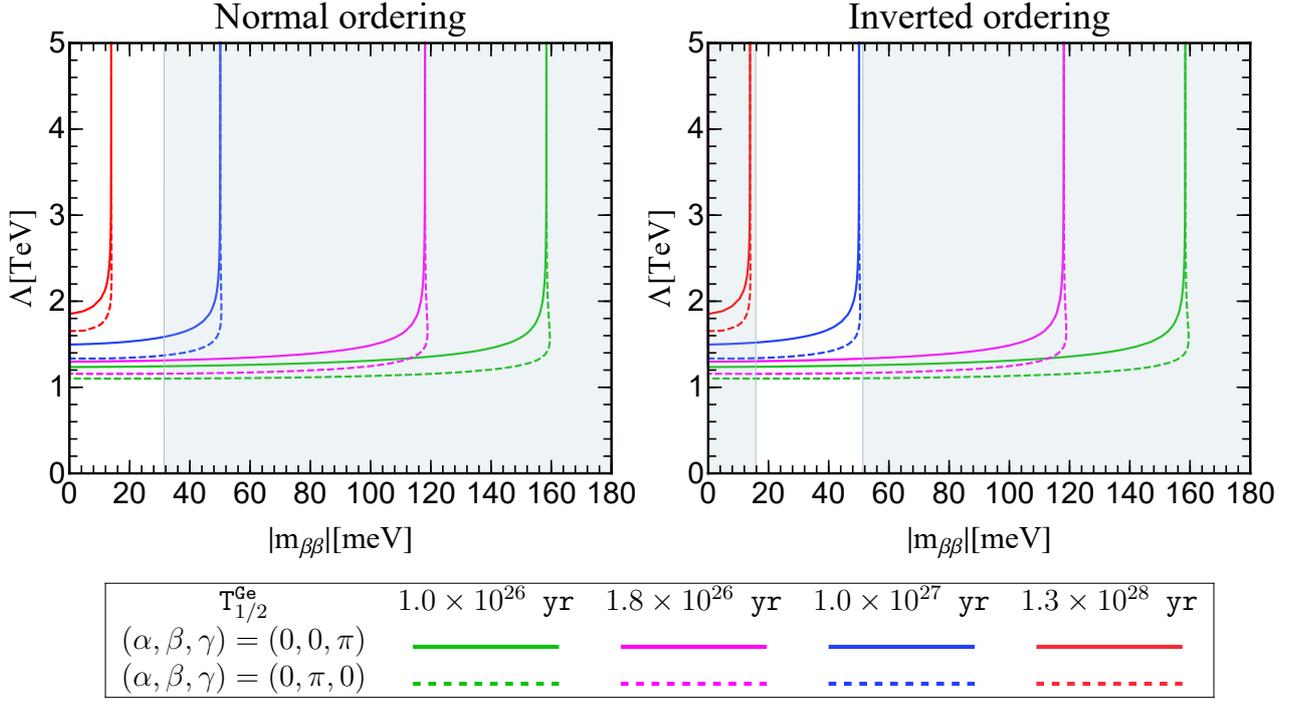

\begin{figure}[t!]
\centering
\includegraphics[scale=0.65]{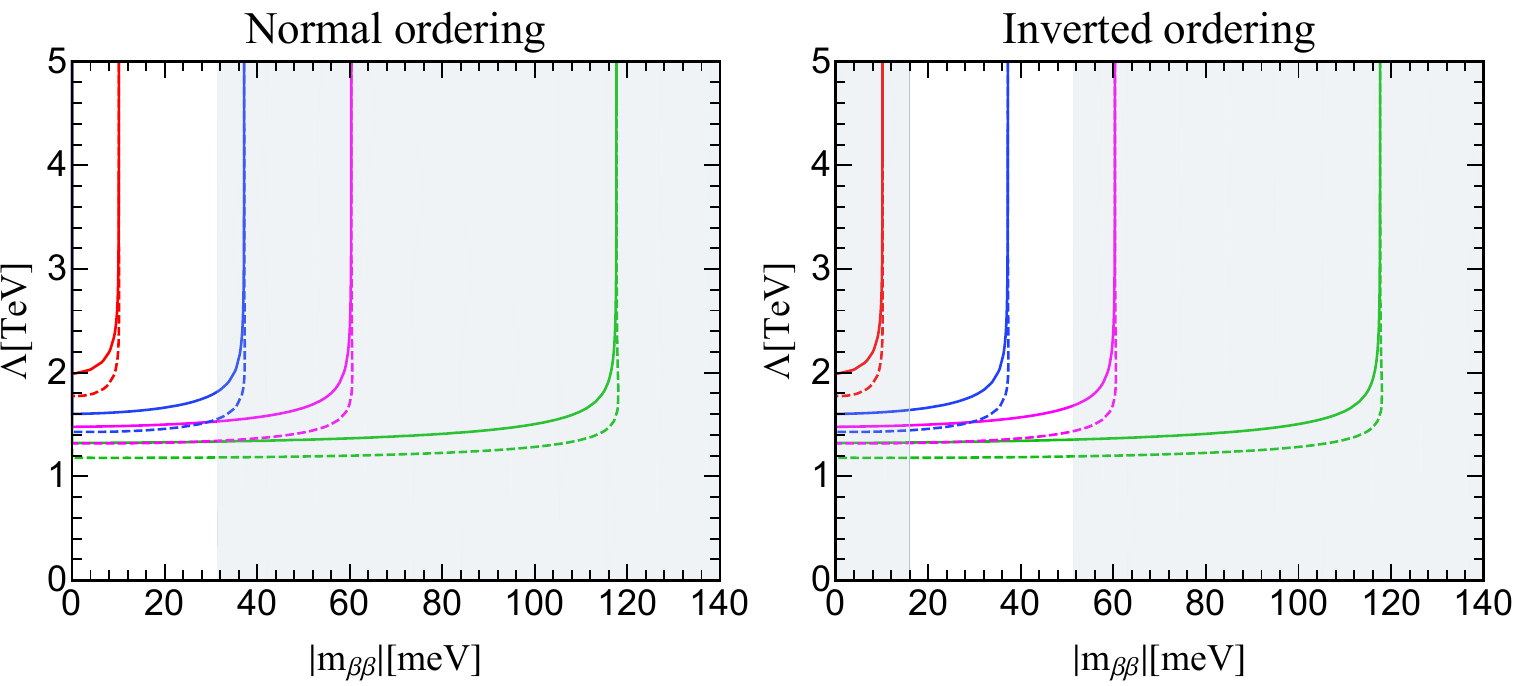}\\[-0.1in]
\qquad\resizebox{0.98\textwidth}{!}{\hspace{1cm}
\begin{tabular}{|ccccc|}\hline
$\texttt{T}_{1/2}^\texttt{Xe}$&$1.0\times10^{26}\texttt{ yr}$&$3.8\times10^{26}\texttt{ yr}$&$1.0\times10^{27}\texttt{yr}$&$1.35\times10^{28}\texttt{ yr}$\\
$(\alpha,\beta,\gamma)=(0,0,\pi)$&\begin{tikzpicture}
\definecolor{greeen}{RGB}{12,194,14};
\draw[greeen,ultra thick] (-1.0,0) --(1.0,0);
\end{tikzpicture}&\begin{tikzpicture}
\definecolor{greeen}{RGB}{255,0,255};
\draw[greeen,ultra thick] (-1.0,0) --(1.0,0);
\end{tikzpicture}&\begin{tikzpicture}
\definecolor{greeen}{RGB}{34,65,250};
\draw[greeen,ultra thick] (-1.0,0) --(1.0,0);
\end{tikzpicture}&\begin{tikzpicture}
\definecolor{greeen}{RGB}{248,42,55};
\draw[greeen,ultra thick] (-1.0,0) --(1.0,0);
\end{tikzpicture}\\
$(\alpha,\beta,\gamma)=(0,\pi,0)$&\begin{tikzpicture}
\definecolor{greeen}{RGB}{12,194,14};
\draw[greeen,ultra thick,dashed] (-1.0,0) --(1.0,0);
\end{tikzpicture}&\begin{tikzpicture}
\definecolor{greeen}{RGB}{255,0,255};
\draw[greeen,ultra thick,dashed] (-1.0,0) --(1.0,0);
\end{tikzpicture}&\begin{tikzpicture}
\definecolor{greeen}{RGB}{34,65,250};
\draw[greeen,ultra thick,dashed] (-1.0,0) --(1.0,0);
\end{tikzpicture}&\begin{tikzpicture}
\definecolor{greeen}{RGB}{248,42,55};
\draw[greeen,ultra thick,dashed] (-1.0,0) --(1.0,0);
\end{tikzpicture}\\\hline
\end{tabular}\hspace{0.6cm}}
\caption{\label{fig:mf3i5-mbbXe-lambda}
Contour plots of the
$0\nu\beta\beta$ decay half-life of $^{136}\text{Xe}$ in the $|m_{\beta\beta}|-\Lambda$ plane for NO and IO neutrino mass spectrums, where $|m_{\beta\beta}|$ is the effective Majorana neutrino mass and $\Lambda$ denotes the mass scale of new fields. The pink and red curves correspond to the current experimental limit $T_{1/2}(^{136}\text{Xe})=3.8\times10^{26}$ yr~\cite{KamLAND-Zen:2024eml} and the projected future sensitivity $T_{1/2}(^{136}\text{Xe})=1.35\times10^{28}$ yr~\cite{nEXO:2021ujk}, respectively. The gray bands denote the regions of $|m_{\beta\beta}|$ disfavored by the current data of neutrino oscillation~\cite{Esteban:2024eli} and the cosmological bound $\sum_{i=1}^3 m_i<0.12$ eV~\cite{Planck:2018vyg}.}
\end{figure}
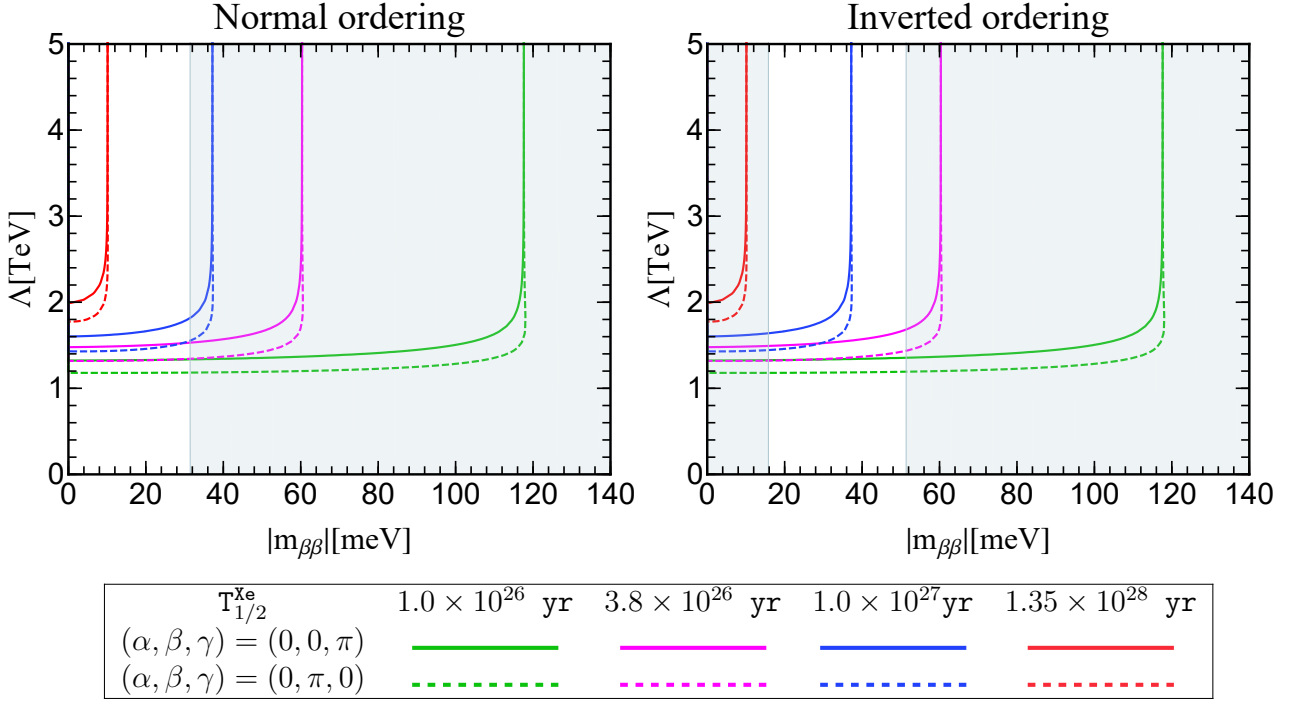

Given the new fields $\phi$, $\zeta$ and $\eta$ in Eq.~\eqref{eq:scalars-model1}, the relevant interactions in Eqs.~(\ref{eq:Lagrangian-model1}, \ref{eq:scalar-poten-model1})
can mediate the can mediate $0\nu\beta\beta$ decay, the corresponding Feynman diagrams are displayed in figure~\ref{fig:model-su3sing-3s}. They are the tree-level UV completions of the dim-11 $0\nu\beta\beta$ operators $\mathcal{O}_{1a}$, $\mathcal{O}_{1b}$, $\mathcal{O}_{2a}$, $\mathcal{O}_{2b}$, $\mathcal{O}_{3a}$ and $\mathcal{O}_{3b}$. After electroweak symmetry breaking, the two external scalars standing for Higgs VEV insertions are removed. The fifteen diagrams in figure~\ref{fig:model_SU3sing_3s_break} in the electroweak basis are reduced to 2 diagrams shown in figure~\ref{fig:model_SU3sing_3s_break} in the mass eigenstate basis. After integrating out all the heavy fields, we obtain the effective Lagrangian for $0\nu\beta\beta$ decay as follow,
\begin{eqnarray}
\mathcal{L}^{\text{eff}}_{0\nu\beta\beta}&=&\Big\{\sum_{i,j,k=1,2}\frac{2f_{i11}f'_{j11}y'^*_{k11}\mu_{ijk}}{m_{\kappa_i}^2m_{\kappa_j}^2m_{\varsigma_k}^2}
+\sum_{i,k=1,2}\frac{ig\left(f'_{i11}m_u-f_{i11}m_d\right)y'^*_{k11}\lambda_{ik}}{\sqrt{2} m_W^2m_{\kappa_i}^2m_{\varsigma_k}^2}\Big\}\left[\overline{u}P_Ld\right]\left[\overline{u}P_Rd\right]\left[\overline{e}P_Le^C\right]\nonumber\\
&&+\Big\{\sum_{i,j,k=1,2}\frac{f_{i11}f_{j11}y'^*_{k11}\mu_{ijk}}{m_{\kappa_i}^2m_{\kappa_j}^2m_{\varsigma_k}^2}
+\sum_{i,k=1,2}\frac{gf_{i11}y'^*_{k11}\lambda_{ik}m_u}{\sqrt{2}m_W^2m_{\kappa_i}^2m_{\varsigma_k}^2}\Big\}
\left[\overline{u}P_Ld\right]\left[\overline{u}P_Ld\right]\left[\overline{e}P_Le^C\right]\nonumber\\
\label{eq:Eff-0nbb-operators-model}&&+\Big\{\sum_{i,j,k=1,2}\frac{f'_{i11}f'_{j11}y'^*_{k11}\mu_{ijk}}{m_{\kappa_i}^2m_{\kappa_j}^2m_{\varsigma_k}^2}
-\sum_{i,k=1,2}\frac{gf'_{i11}y'^*_{k11}\lambda_{ik}m_d}{\sqrt{2}m_W^2m_{\kappa_i}^2m_{\varsigma_k}^2}\Big\}
\left[\overline{u}P_Rd\right]\left[\overline{u}P_Rd\right]\left[\overline{e}P_Le^C\right]\,.
\end{eqnarray}
In terms of the basis operator of $0\nu\beta\beta$ decay at low energy scale in appendix~\ref{sec:0nubb-eff-operators-LEFT}, the above effective Lagrangian can be written as
\begin{eqnarray}
\mathcal{L}^{\text{eff}}_{0\nu\beta\beta}&=&\frac{G^2_F\cos^2\theta_C}{2m_P}\left[\epsilon_{1LL}^R\left(\mathcal{O}_1\right)_{\left\{LR\right\}R}+
\epsilon_{1RR}^R\left(\mathcal{O}_1\right)_{\left\{LL\right\}R}+\epsilon_{1LR}^R\left(\mathcal{O}_{1}\right)_{\left\{RR\right\}R}
\right]\,,
\end{eqnarray}
where
\begin{eqnarray}
\epsilon_{1LL}^R&=&\frac{m_P}{G^2_F\cos^2\theta_C}\Big\{\sum_{i,j,k=1,2}\frac{f_{i11}f_{j11}y'^*_{k11}\mu_{ijk}}{4m_{\kappa_i}^2m_{\kappa_j}^2m_{\varsigma_k}^2}
+\sum_{i,k=1,2}\frac{gf_{i11}y'^*_{k11}\lambda_{ik}m_u}{4\sqrt{2}m_W^2m_{\kappa_i}^2m_{\varsigma_k}^2}\Big\}\nonumber\\
\epsilon_{1RR}^R&=&\frac{m_P}{G^2_F\cos^2\theta_C}\Big\{\sum_{i,j,k=1,2}\frac{f'_{i11}f'_{j11}y'^*_{k11}\mu_{ijk}}{4m_{\kappa_i}^2m_{\kappa_j}^2m_{\varsigma_k}^2}
-\sum_{i,k=1,2}\frac{gf'_{i11}y'^*_{k11}\lambda_{ik}m_d}{4\sqrt{2}m_W^2m_{\kappa_i}^2m_{\varsigma_k}^2}\Big\}\nonumber\\
\label{eq:epsilon-coeffs}\epsilon_{1LR}^R&=&\frac{m_P}{G^2_F\cos^2\theta_C}\Big\{\sum_{i,j,k=1,2}\frac{f_{i11}f'_{j11}y'^*_{k11}\mu_{ijk}}{2m_{\kappa_i}^2m_{\kappa_j}^2m_{\varsigma_k}^2}
+\sum_{i,k=1,2}\frac{g\left(f'_{i11}m_u-f_{i11}m_d\right)y'^*_{k11}\lambda_{ik}}{4\sqrt{2}m_W^2m_{\kappa_i}^2m_{\varsigma_k}^2}\Big\}\,.
\end{eqnarray}
where $g$ is the coupling constant of $SU(2)_L$. We see that our example model involves three $0\nu\beta\beta$ decay operators $\left(\mathcal{O}_1\right)_{\left\{LR\right\}R}$, $\left(\mathcal{O}_1\right)_{\left\{LL\right\}R}$ and $\left(\mathcal{O}_{1}\right)_{\left\{RR\right\}R}$ at low energy. As shown in appendix~\ref{sec:0nubb-eff-operators-LEFT}, the half-life of $0\nu\beta\beta$ decay can be expressed in terms of the Wilson coefficients of the low energy $0\nu\beta\beta$ decay operators, the phase-space factors (PSFs) and nuclear matrix elements (NMEs). Using the general formula Eq.~\eqref{eq:inverse-half-life}, we find that the inverse half-life of $0\nu\beta\beta$ in this model is determined to be
\begin{eqnarray}
T_{1/2}^{-1}&=&G_{11+}^{(0)}\left|\epsilon_\nu\mathcal{M}_\nu\right|^2+G_{11+}^{(0)}\left|\epsilon_{1LL}^R\mathcal{M}_{1LL}+\epsilon_{1RR}^R\mathcal{M}_{1RR}+\epsilon_{1LR}^R\mathcal{M}_{1LR}\right|^2\nonumber\\
&&+2G_{11-}^{(0)}\Re\Big[\epsilon_\nu\mathcal{M}_\nu\left(\epsilon_{1LL}^R\mathcal{M}_{1LL}+\epsilon_{1RR}^R\mathcal{M}_{1RR}+\epsilon_{1LR}^R\mathcal{M}_{1LR}\right)^*\Big]\,,
\end{eqnarray}
where $G_{11+}^{(0)}$, $G_{11-}^{(0)}$ and $G_{11+}^{(0)}$ are the phase space factors, and $\mathcal{M}_\nu$, $\mathcal{M}_{1LL}$, $\mathcal{M}_{1RR}$ and $\mathcal{M}_{1LR}$ are the nuclear matrix element parts of the $0\nu\beta\beta$ decay amplitude. Their definitions are provided in appendix~\ref{sec:0nubb-eff-operators-LEFT}. When the phases of the Wilson coefficients are taken into account, the half-life of neutrinoless double beta decay takes the following form:
\begin{eqnarray}
T_{1/2}^{-1}&=&G_{11+}^{(0)}\left|\epsilon_\nu\mathcal{M}_\nu\right|^2+G_{11+}^{(0)}\Big|\left|\epsilon_{1LR}^R\mathcal{M}_{1LR}\right|
+e^{i\alpha}\left|\epsilon_{1LL}^R\mathcal{M}_{1LL}\right|+e^{i\beta}\left|\epsilon_{1RR}^R\mathcal{M}_{1RR}\right|\Big|^2\nonumber\\
&&+2G_{11-}^{(0)}\Re\Big[\left|\epsilon_\nu\mathcal{M}_\nu\right|e^{i\gamma}\left(\left|\epsilon_{1LR}^R\mathcal{M}_{1LR}\right|
+e^{i\alpha}\left|\epsilon_{1LL}^R\mathcal{M}_{1LL}\right|+e^{i\beta}\left|\epsilon_{1RR}^R\mathcal{M}_{1RR}\right|\right)^*\Big]\nonumber\\
&=&G_{11+}^{(0)}\left|\epsilon_\nu\mathcal{M}_\nu\right|^2+G_{11+}^{(0)}\left|\epsilon_{1LR}^R\mathcal{M}_{1LR}\right|^2
+G_{11+}^{(0)}\left|\epsilon_{1LL}^R\mathcal{M}_{1LL}\right|^2+G_{11+}^{(0)}\left|\epsilon_{1LL}^R\mathcal{M}_{1LL}\right|^2\,,\nonumber\\
&&+2G_{11+}^{(0)}\left|\epsilon_{1LR}^R\mathcal{M}_{1LR}\right|\left|\epsilon_{1LL}^R\mathcal{M}_{1LL}\right|\cos\alpha
+2G_{11+}^{(0)}\left|\epsilon_{1LR}^R\mathcal{M}_{1LR}\right|\left|\epsilon_{1RR}^R\mathcal{M}_{1RR}\right|\cos\beta\nonumber\\
&&+2G_{11+}^{(0)}\left|\epsilon_{1LL}^R\mathcal{M}_{1LL}\right|\left|\epsilon_{1RR}^R\mathcal{M}_{1RR}\right|\cos\left(\alpha-\beta\right)
+2G_{11-}^{(0)}\left|\epsilon_\nu\mathcal{M}_\nu\right|\left|\epsilon_{1LR}^R\mathcal{M}_{1LR}\right|\cos\gamma\nonumber\\
&&+2G_{11-}^{(0)}\left|\epsilon_\nu\mathcal{M}_\nu\right|\left|\epsilon_{1LL}^R\mathcal{M}_{1LL}\right|\cos\left(\gamma-\alpha\right)
+2G_{11-}^{(0)}\left|\epsilon_\nu\mathcal{M}_\nu\right|\left|\epsilon_{1RR}^R\mathcal{M}_{1RR}\right|\cos\left(\gamma-\beta\right)\,.~~~~\label{eq:inv-halflife}
\end{eqnarray}
For illustration, we present contour plots of the half-life of $^{76}\text{Ge}$ and $^{136}\text{Xe}$ in the plane of $\Lambda$ versus $|m_{\beta\beta}|$ in figures~\ref{fig:mf3i5-mbbGe-lambda} and \ref{fig:mf3i5-mbbXe-lambda}. Here $|m_{\beta\beta}|$ is the effective Majorana neutrino mass defined as $|m_{\beta\beta}|=|\sum_{i=1}^3 U_{ei}^2 m_i|$, where $m_i$ denote the light neutrino masses and $U_{ei}$ are elements of the lepton mixing matrix. In our numerical analysis, all masses of the new fields are taken to be equal, $m_\phi=m_\zeta=m_\eta=\Lambda$. The other relevant model parameters are fixed to
\begin{align}
\nonumber &y_{\alpha\beta}=h''_{\alpha\beta}=0.1\,,\qquad h_{\alpha\beta}=h'_{\alpha\beta}=1\,,\\
&\lambda_{13}=\lambda_{14}=\lambda_{17}=0.01\,,\qquad\mu_1=\mu_2=100~\text{GeV}\,,
\label{eq:para-choice}
\end{align}
unless stated otherwise. When plotting figures~\ref{fig:mf3i5-mbbGe-lambda} and \ref{fig:mf3i5-mbbXe-lambda}, we have adopted representative values of the phase-space factors (PSFs) and nuclear matrix elements (NMEs) in the interacting boson model IBM-2 from Ref.~\cite{Deppisch:2020ztt}. Using the latest global-fit results of neutrino oscillation parameters~\cite{Esteban:2024eli} and the bound on the light neutrino mass $\sum_{i=1}^3 m_i<0.12$ from Planck~\cite{Planck:2018vyg}, the allowed ranges of the effective Majorana mass are found to be $|m_{\beta\beta}|\leq 31.42~\text{meV}$ for normal ordering (NO), and $15.80~\text{meV}\leq |m_{\beta\beta}|\leq 51.48~\text{meV}$ for inverted ordering (IO). The regions of $|m_{\beta\beta}|$ disfavored by the current neutrino oscillation and Cosmological data are indicated by gray bands in these figures. we see that the present bounds $T_{1/2}(^{76}\text{Ge})>1.8\times10^{26}$ yr~\cite{GERDA:2020xhi} and $T_{1/2}(^{136}\text{Xe})>3.8\times10^{26}$ yr~\cite{KamLAND-Zen:2024eml} can be accommodated in a large parameter space for both NO and IO. However, for the sensitivity $T_{1/2}>10^{28}$ yr which is expected to be achieved for the isotopes Xe or Ge in future, the IO region is expected to be excluded by $0\nu\beta\beta$ decay experiments even in the presence of exotic short-range contributions. In this case, the allowed region of $|m_{\beta\beta}|$ would be restricted to a narrow interval for NO, thereby leading to strong constraints on the neutrino masses and mixing parameters.

From Eq.~\eqref{eq:epsilon-coeffs}, we can see that the Wilson coefficients $\epsilon_{1LL}^R$, $\epsilon_{1RR}^R$, $\epsilon_{1LR}^R$ scale as $\Lambda^{-6}$. Therefore the short-range contributions to the inverse half-life $T^{-1}_{1/2}$ is proportional to $\Lambda^{-12}$. For the parameter choice in Eq.~\eqref{eq:para-choice}, the short-range contribution dominates over the mass mechanism if the new physics scale $\Lambda$ is less than few TeV, and then variations of relative phases $\alpha$, $\beta$, and $\gamma$ would have a noticeable impact on the half-life. On the other hand, the $0\nu\beta\beta$ decay rate would be dominantly governed by the long-range contribution and its dependence on $m_{\beta\beta}$ for larger new physics scale $\Lambda$.

\subsection{Generation of neutrino masses}

\begin{figure}[htbp]
\centering
\includegraphics[width=0.90\textwidth]{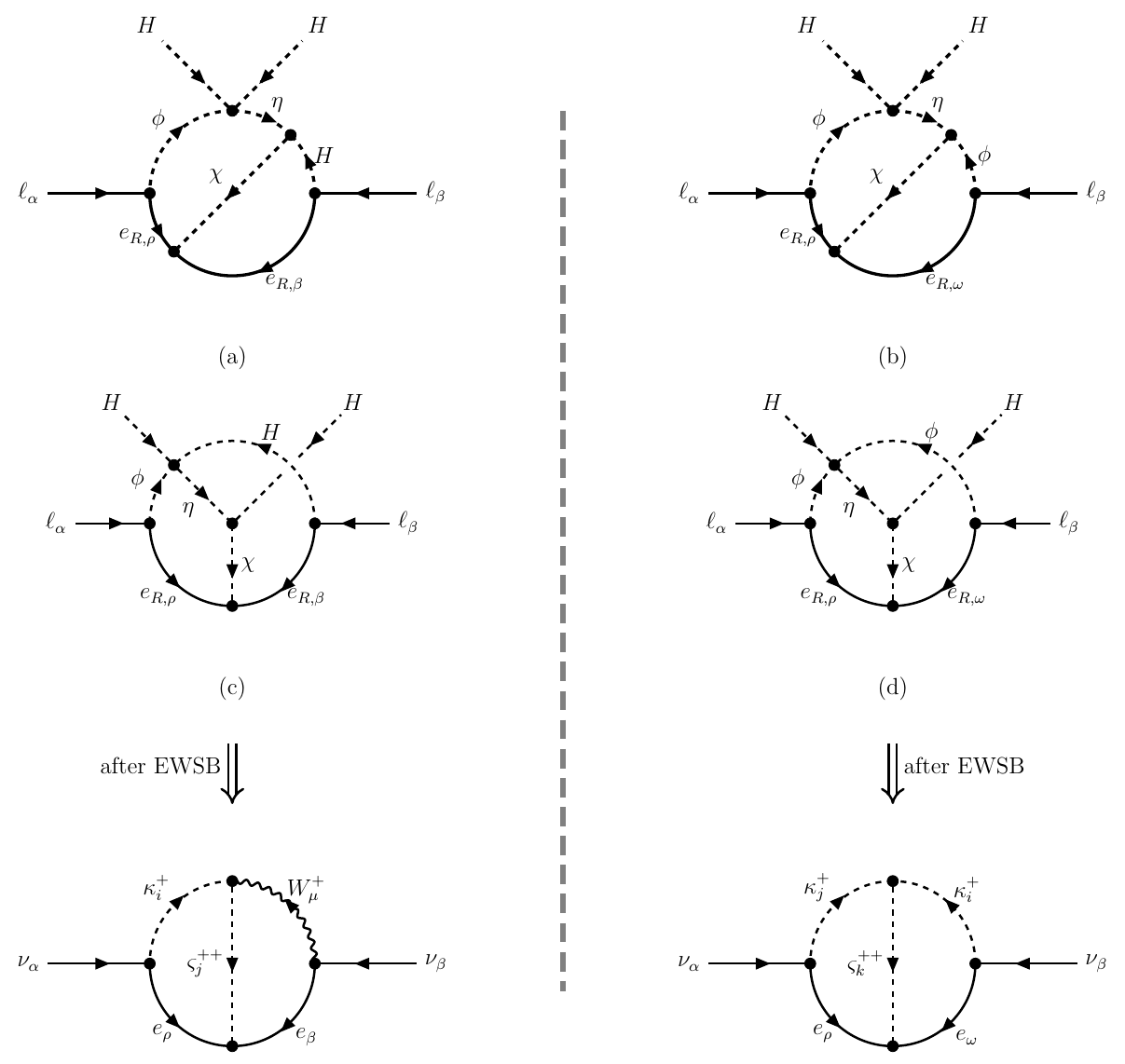}
\caption{\label{fig:model-su3sing-3s-wein-2l}Neutrino mass generation
in the representative $0\nu\beta\beta$ model. }
\end{figure}

In $0\nu\beta\beta$ decay, the lepton number is violated by two units. Consequently any ultraviolet completion of the corresponding effective operator must include lepton number violating interactions, which can induce Majorana neutrino masses. In this model, there are four Feynman diagrams  contributing to light neutrino masses in the electroweak basis, and they reduce to two independent diagrams in the mass basis, as shown in figure~\ref{fig:model-su3sing-3s-wein-2l}. The resulting neutrino mass matrix can be computed straightforwardly as:
\begin{eqnarray}
(M_\nu)_{\alpha\beta}&=&\sum_{\substack{i,j=1,2\\
\rho=1,2,3}}\Big\{\frac{\sqrt{2}gf''^*_{i\alpha\rho}y'_{j\beta\rho}\lambda_{ij}m_{e_\beta}}{\left(2\pi\right)^8}\Big[\hat{\mathcal{I}}^{p\cdot q}\left(m_{e_\beta}^2,m_W^2,m_{e_\rho}^2,m_{\kappa_i}^2,m_{\varsigma_j}^2\right)\nonumber\\
&&-\frac{1}{m_W^2}\hat{\mathcal{I}}^{p^2(p\cdot q)}\left(m_{e_\beta}^2,m_W^2,m_{e_\rho}^2,m_{\kappa_i}^2,m_{\varsigma_j}^2\right)\Big]+\left(\alpha\leftrightarrow\beta\right)\Big\}\nonumber\\
&&-\sum_{\substack{i,j,k=1,2\\
\rho,\omega=1,2,3}}\frac{4f''^*_{j\alpha\rho}y'_{k\rho\omega}f''^*_{i\beta\omega}\mu^*_{ijk}}{\left(2\pi\right)^8}\hat{\mathcal{I}}^{p\cdot q}\left(m_{e_\omega}^2,m_{\kappa_i}^2,m_{e_\rho}^2,m_{\kappa_j}^2,m_{\varsigma_k}^2\right)\,, \label{eq:mnu-alpha-beta}
\end{eqnarray}
where $\alpha$, $\beta$, $\rho$ and $\omega$ are flavor indices, the parameters $f''_{j\alpha\rho}$, $y'_{i\beta\rho}$, $\lambda_{ij}$, and $\mu_{ijk}$ are defined in Eqs.~\eqref{eq:f-h-fp-fpp-yp-def}, \eqref{eq:lambda-ij}, and \eqref{eq:mu-ijk}, respectively. They are expressed in terms of the fundamental couplings of the model and the mixing angles $\theta_\kappa$ and $\theta_\varsigma$ that relate the interaction eigenstates to the mass eigenstates. The analytical expressions of the two-loop integrals $\hat{\mathcal{I}}^{p\cdot q}$ and $\hat{\mathcal{I}}^{p^2(p\cdot q)}$ are provided in the appendix~\ref{app:2loop-def}. Although both $\hat{\mathcal{I}}^{p\cdot q}$ and $\hat{\mathcal{I}}^{p^2(p\cdot q)}$ are individually divergent, the divergences cancel exactly in each matrix element $(M_\nu)_{\alpha\beta}$.

The first two lines of $(M_{\nu})_{\alpha\beta}$ in Eq.~\eqref{eq:mnu-alpha-beta} are generated by the Feynman diagrams shown on the left of figure~\ref{fig:model-su3sing-3s-wein-2l} and scale with the charged-lepton masses $m_{\alpha}$ or $m_{\beta}$. In contrast, the contribution in the last line of Eq.~\eqref{eq:mnu-alpha-beta} arises from the diagrams displayed on the right side of figure~\ref{fig:model-su3sing-3s-wein-2l}. It depends on the mass parameters $\mu_{ijk}$, which are linear combinations of $\mu_2$ and the Higgs VEV $v$, as can be seen from Eq.~\eqref{eq:mu-ijk}. The light neutrino mass also depends on another mass parameter $\mu_1$
through the mass eigenvalues $m_{\varsigma_1}$, $m_{\varsigma_2}$ and the scalar mixing angle $\theta_\varsigma$. Since the charged-lepton masses are typically smaller than $\mu_{ijk}$ in the absence of fine-tuning or cancellations, the light neutrino mass is dominantly determined by the last term of Eq.~\eqref{eq:mnu-alpha-beta}. The light neutrino mass would vanish in the degenerate limit $m_{\varsigma_1}=m_{\varsigma_2}$ and $m_{\kappa_1}=m_{\kappa_2}$. Therefore the mass splittings between $\varsigma^{++}_1$ and $\varsigma^{++}_2$, as well as between $\kappa^+_1$ and $\kappa^+_2$, are responsible for generating small neutrino masses.

\begin{figure}[hptb]
\centering
\includegraphics[scale=0.29]{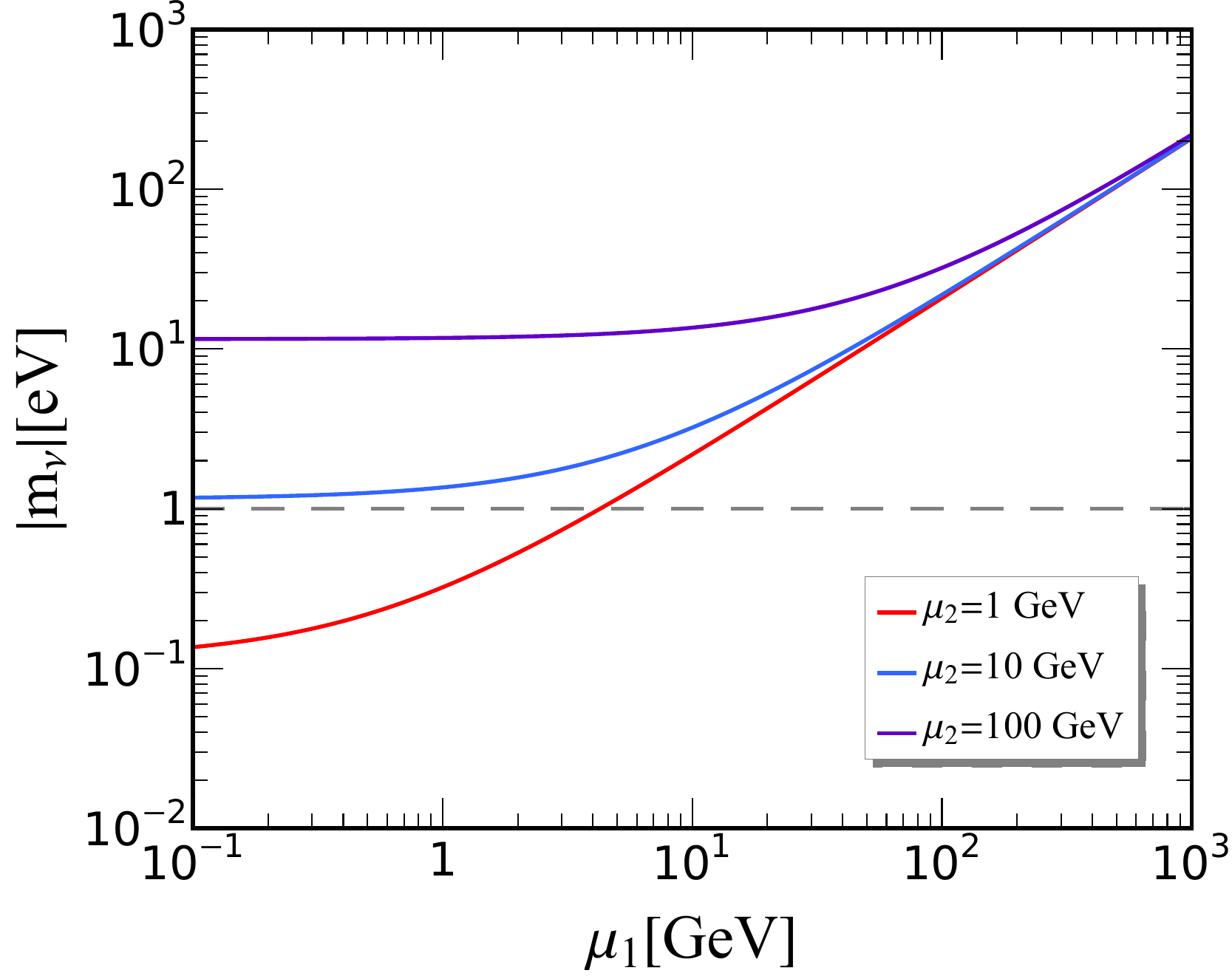}
\includegraphics[scale=0.29]{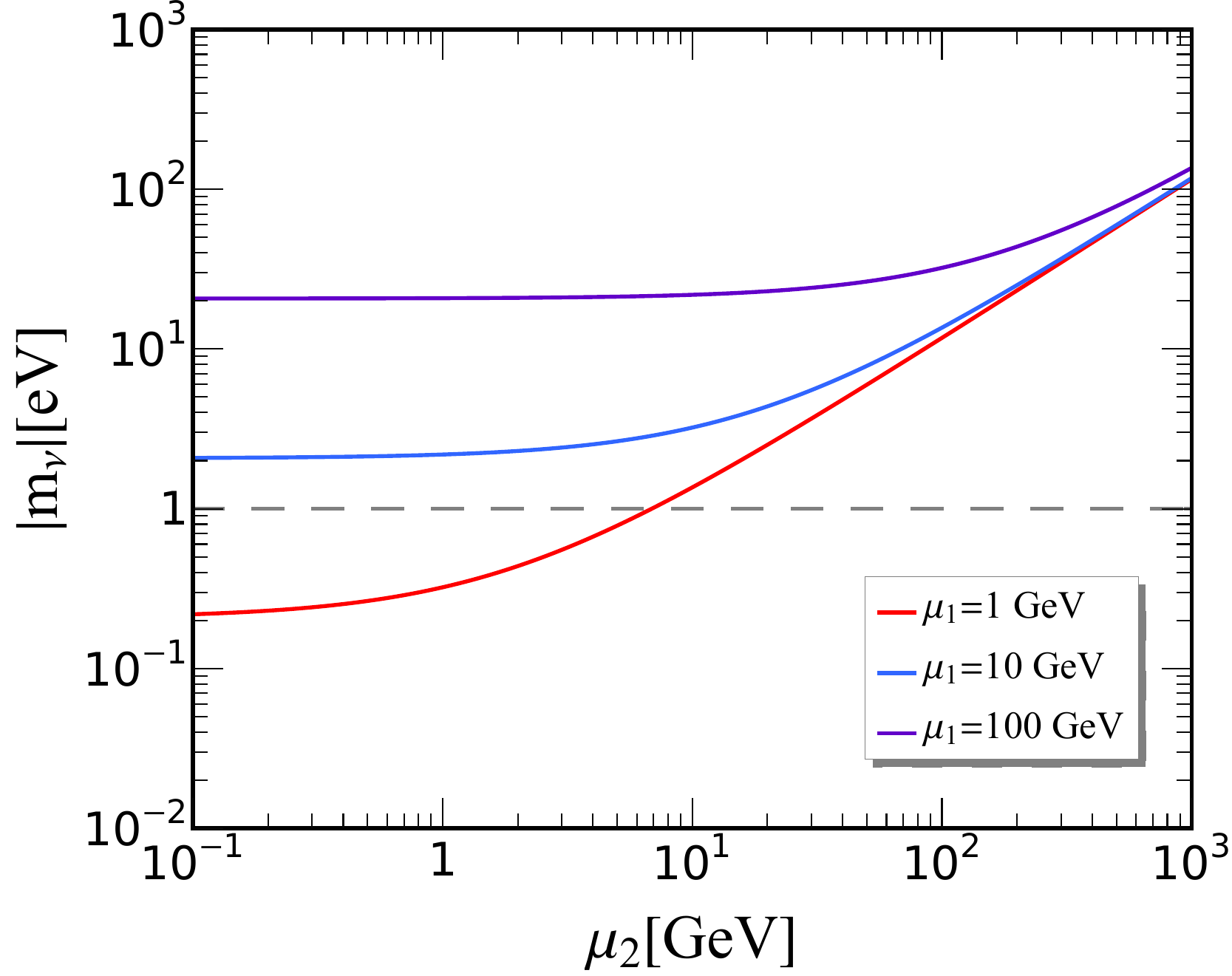}
\caption{The light neutrino mass as a function of $\mu_1$( left panel) and $\mu_2$( right panel), where we take the mediator mass $m_\zeta=m_\eta=m_\phi=1$ TeV for illustration. \label{fig:mnu-vs-mu12}}
\end{figure}

We show the variation of neutrino mass with respect to $\mu_1$ and $\mu_2$ in figure~\ref{fig:mnu-vs-mu12}, where we take $\lambda_6=\lambda_7=\lambda_9=\lambda_{12}=0.01$ and all remaining parameters are taken to be the same as those specified in Eq.~\eqref{eq:para-choice}, and the three new scalars $\eta$, $\phi$ and $\zeta$ are assumed to be of the same masses with $m_{\eta}=m_{\phi}=m_{\zeta}\equiv \Lambda=1$ TeV. In our numerical analysis, we ignore any flavour structures in the Yukawa sector and take all Yukawa couplings to be flavour-universal, as shown in Eq.~\eqref{eq:para-choice}. Therefore the numerical values of $|m_{\nu}|$ should be viewed as order of magnitude estimate of the neutrino mass scale, not as exact predictions. From figure~\ref{fig:mnu-vs-mu12} we see that the observed neutrino mass scale can be obtained for values of $\mu_1$ and $\mu_2$ as low as a few GeV, provided the new physics scale is of order TeV.

\begin{figure}[t!]
\centering
\includegraphics[scale=0.40]{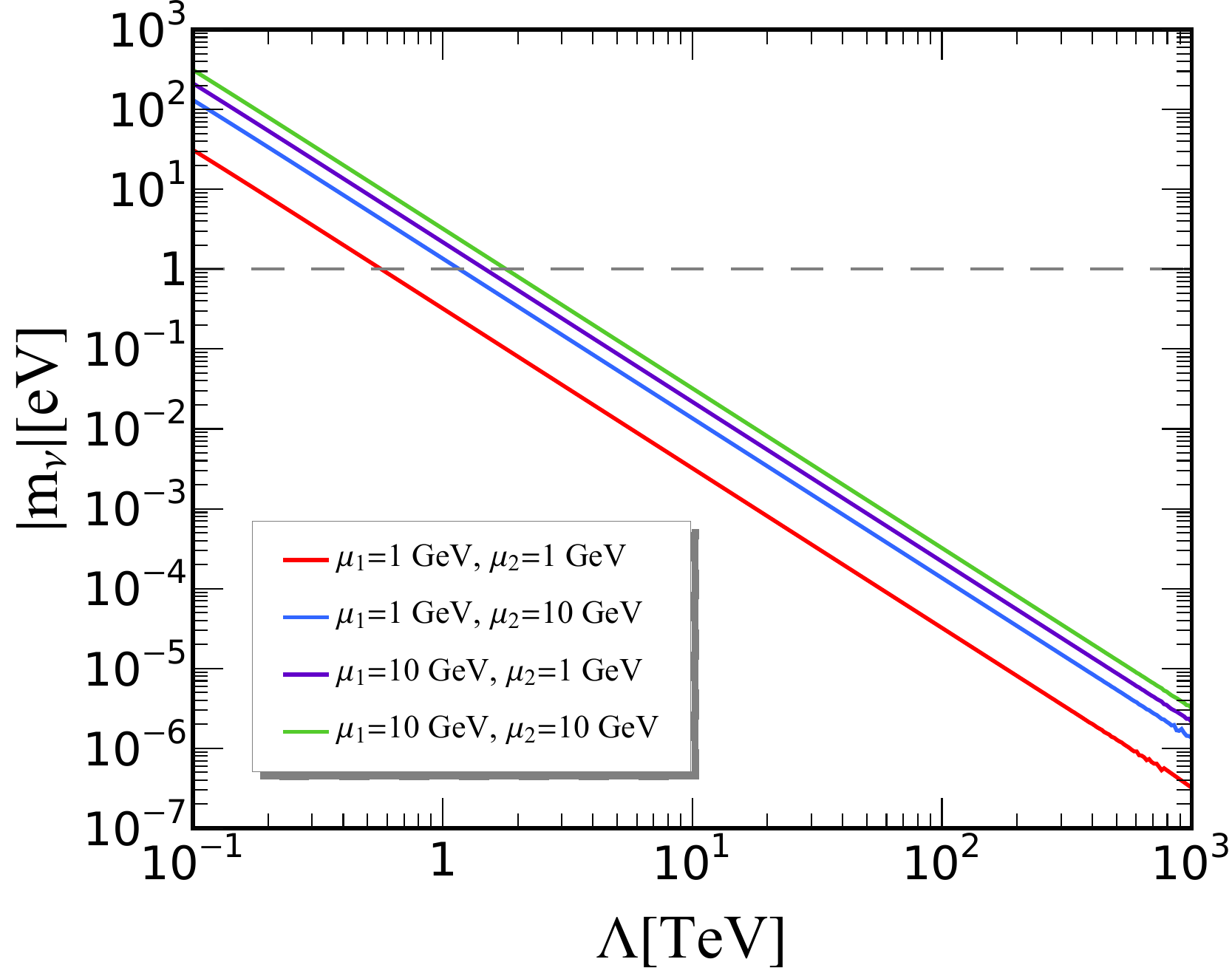}
\caption{The light neutrino mass with respect to mediator mass $\Lambda$ for some typical choices of the values of $\mu_1$ and $\mu_2$, where we take all masses of the new scalar equal with $m_\zeta=m_\eta=m_\phi=\Lambda$. \label{fig:mnu-vs-Lambda} }
\end{figure}

Furthermore, we show the dependence of the neutrino mass on the masses of the mediators in figure~\ref{fig:mnu-vs-Lambda}, where we take $m_\zeta=m_\eta=m_\phi=\Lambda$. The mass splittings in the $\varsigma^{++}_{1,2}$ and $\kappa^+_{1,2}$ sectors decrease as $\Lambda$ increases. Consequently the neutrino mass $M_\nu$ decreases with increasing $\Lambda$ and it tends to zero in the decoupling limit $\Lambda \to \infty$, as shown in figure~\ref{fig:mnu-vs-Lambda}. The correct neutrino mass scale can be reproduced for mediator masses $\Lambda \sim \mathcal{O}(1)~\mathrm{TeV}$ when the Yukawa couplings are taken to be of order $\mathcal{O}(0.1)$. We remark that the neutrino mass matrix in Eq.~\eqref{eq:mnu-alpha-beta} depends on the Yukawa couplings $f''_{i\alpha\beta}$ and $y'_{i\alpha\beta}$, which are $3\times3$ complex matrices in flavour space. As a result, the model contains sufficient free parameters to accommodate all current neutrino oscillation data, including the three mixing angles and the two independent neutrino mass-squared differences.

\section{\label{sec:conclusion}Conclusion}

The $0\nu\beta\beta$ decay, which violates lepton number by two units, is a sensitive probe of the Majorana nature of neutrinos and offers valuable information on the neutrino mass spectrum and mixing parameters. The standard interpretation assumes that $0\nu\beta\beta$ decay is dominated by the exchange of light Majorana neutrinos. However, a wide class of scenarios beyond the SM can also induce $0\nu\beta\beta$ decay. Such contributions are typically categorized into long-range and short-range mechanisms, mediated by light neutrinos and heavy degrees of freedom, respectively. At low energies, the effective Lagrangian describing
$0\nu\beta\beta$ decay can be written in terms of products of leptonic and hadronic currents. The ultraviolet completions of the corresponding effective operators allow for a systematic exploration of the possible origins of $0\nu\beta\beta$ decay.

The effective Lagrangian for short-range $0\nu\beta\beta$ decay comprises twenty-four independent operators with distinct Lorentz structures. Eleven of these operators originate from dim-9 operators which are invariant under the SM gauge symmetry. Among the remaining thirteen operators, twelve arise from dim-11 SM invariant operators, while the last one first appears at dimension thirteen. Both tree and one-loop decompositions of the dim-9 short-range $0\nu\beta\beta$ operators have been studied in Refs.~\cite{Bonnet:2012kh,Chen:2021rcv}. The present work provides a systematic analysis of the tree-level decompositions of the dim-11 $0\nu\beta\beta$ decay operators, identifying all possible ultraviolet completions and the corresponding mediators. We restrict our analysis to fermionic and scalar mediators, since introducing vector mediators beyond the SM gauge bosons would necessitate an extension of the SM gauge symmetry. Nevertheless, many of our results obtained for scalar mediators can be straightforwardly extended to scenarios involving vector mediators.

The ultraviolet completions are assumed to be renormalizable above the electroweak scale, consequently the allowed interaction vertices are restricted to scalar–scalar–scalar, fermion–fermion–scalar, and scalar–scalar–scalar–scalar forms. The dim-11 $0\nu\beta\beta$ operators contain four quark fields, two lepton fields and two Higgs fields. We find eight distinct topologies for the tree-level decomposition of these operators, as shown in figure~\ref{fig:topology}. By specifying the Lorentz nature (scalar or fermion) of each line, the topologies can be promoted to diagrams. This yields twenty-eight possible diagrams with two external scalars and six external fermion fields, shown in figure~\ref{fig:genuine-diagrams}. Subsequently, the quark, lepton, and Higgs fields appearing in the dim-11 $0\nu\beta\beta$ operators are assigned to the external legs. The SM gauge quantum numbers of all internal fields are then uniquely determined by requiring gauge invariance at each interaction vertex. This procedure allows for a systematic generation of all possible UV completions. Our analysis is restricted to models in which the leading contribution to $0\nu\beta\beta$ decay arises from dim-11 operators. This condition excludes scenarios where the mediators generate dim-9 short-range operators at tree level. Moreover, we impose that light Majorana neutrino masses are induced no earlier than at two loops, so as to prevent the suppression of the dim-11 contribution.

We identify a large class of models realizing the dim-11 $0\nu\beta\beta$ decay operators, which are collected in the supplementary file~\cite{Li:2026dim11NDBDsup}. In total, these constructions involve 61 new fields beyond the SM, which are summarized in table~\ref{tab:allnf}, with individual models involving three, four or five mediators. They generically feature fractionally charged fermions and exotic bosons such as dileptons, diquarks and leptoquarks. Consequently a rich collider phenomenology is expected. Lepton number violating interactions are generically present and provide an important probe of these models.

Nearly all ultraviolet completions involve colored mediator fields, with the exception of two models presented in section~\ref{sec:representative-model} and appendix~\ref{app:add-onbb-models}, respectively. The first model requires three colorless scalar fields: a diquark doublet $\phi$, a dilepton singlet $\zeta$, and a scalar doublet $\eta$ carrying exotic electric charges. The second model differs in that the dilepton scalar $\zeta$ is replaced by an uncolored vector-like fermion doublet $\Psi$. For the first model, we investigate the implications for the $0\nu\beta\beta$ decay half-life and the generation of light neutrino masses. We find that forthcoming ton-scale $0\nu\beta\beta$ decay experiments place stringent constraints on the allowed parameter space. The observed small neutrino masses can be achieved for mediator masses in the range of few TeV, which is partially accessible at current and future colliders, as well as in experiments searching for charged lepton flavour violation. The systematic decomposition of dim-11 $0\nu\beta\beta$ operators carried out in this work is expected to provide a useful guide for identifying and interpreting exotic short-range contributions to $0\nu\beta\beta$ decay if a signal is observed in forthcoming experiments.

\section*{Acknowledgements}

SYL and GJD are supported by the National Natural Science Foundation of China under Grant Nos.~12375104, 12547106 and Guizhou Provincial Major Scientific and Technological Program XKBF (2025)010.

\clearpage

\appendix

\section{\label{app:non-renor-top}Non-renormalizable topologies of dim-11 $0\nu\beta\beta$ decay operators at tree-level }

\begin{figure}[htbp]
\centering
\includegraphics[scale=0.4]{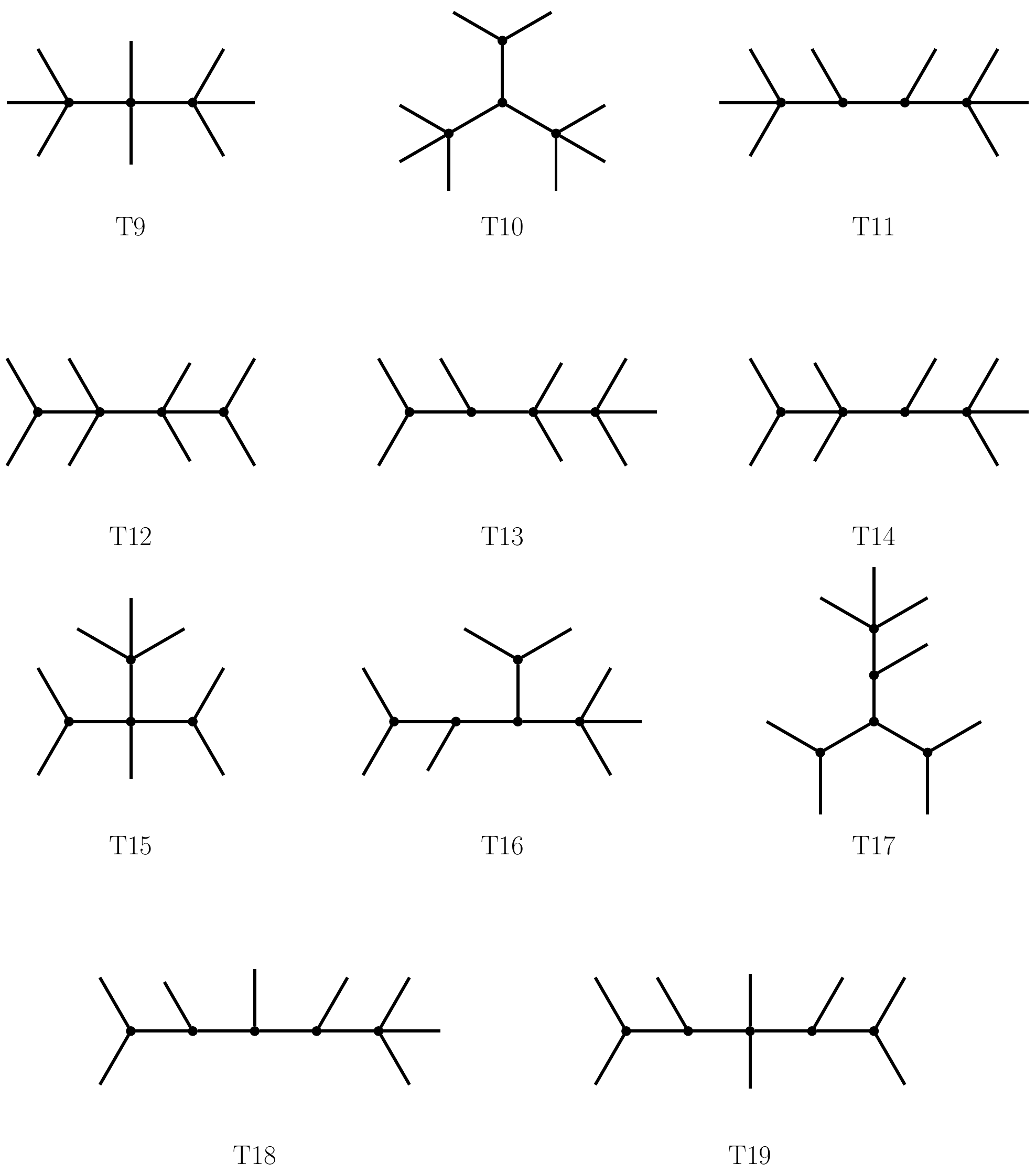}
\caption{\label{fig:non-renormalizable-topologies}The non-renormalizable topologies for the tree-level UV completions of the dim-11 $0\nu\beta\beta$ operators. }
\end{figure}

In addition to the eight topologies shown in figure~\ref{fig:topology}, there are eleven further tree-level topologies with eight external lines, displayed in figure~\ref{fig:non-renormalizable-topologies}. As can be seen, the dim-11
$0\nu\beta\beta$ decay operators in Eq.~\eqref{eq:0nbb-d11-opers} contain six fermion fields and two scalar fields. When quark, lepton, and Higgs fields are attached to the external lines, some interaction vertices necessarily involve either two fermions and two scalars or four fermions, since Lorentz invariance requires each interaction term in the Lagrangian to contain an even number of fermion fields. The corresponding interaction terms therefore have mass dimension greater than four and are non-renormalizable by power counting. In this work, we focus on renormalizable ultraviolet (UV) completions of the dim-11 $0\nu\beta\beta$ decay operators at tree level.

\section{\label{sec:0nubb-eff-operators-LEFT} Low energy effective operators and half-life of short-range $0\nu\beta\beta$ decay }

After electroweak symmetry breaking, the SM gauge symmetry $SU(3)_C\times SU(2)_L\times U(1)_Y$ is reduced to $SU(3)_C\times U(1)_{EM}$. At short distances, the leading contributions to $0\nu\beta\beta$ decay originate from operators containing four quark fields and two charged leptons. These operators are invariant under $SU(3)_{C}\times U(1)_{EM}$ and they can be expressed as products of three fermion currents~~\cite{Pas:2000vn,Deppisch:2020ztt},
\begin{eqnarray}
\nonumber&&(\mathcal{O}_{1})_{\{LL\}L}= J_{L}J_{L}j_{L}\,,\quad~~~ (\mathcal{O}_{1})_{\{LL\}R}=J_{L}J_{L}j_{R}\,,\quad~~~(\mathcal{O}_{1})_{\{RR\}L}= J_{R}J_{R}j_{L}\,,\\
\nonumber&&(\mathcal{O}_{1})_{\{RR\}R}=J_{R}J_{R}j_{R}\,,\quad~~~(\mathcal{O}_{1})_{\{LR\}L}= J_{L}J_{R}j_{L}\,,\quad~~~(\mathcal{O}_{1})_{\{LR\}R}=J_{L}J_{R}j_{R}\,,\\
\nonumber&&(\mathcal{O}_{2})_{\{LL\}L}=J^{\mu\nu}_{L}J_{L\mu\nu}j_{L}\,,\quad~~~
(\mathcal{O}_{2})_{\{LL\}R} =J^{\mu\nu}_{L}J_{L\mu\nu}j_{R}\,,\quad~~~ (\mathcal{O}_{2})_{\{RR\}L}=J^{\mu\nu}_{R}J_{R\mu\nu}j_{L}\,,\\
\nonumber&&(\mathcal{O}_{2})_{\{RR\}R}=J^{\mu\nu}_{R}J_{R\mu\nu}j_{R}\,,\quad~~~(\mathcal{O}_{3})_{\{LL\}L}= J^{\mu}_{L}J_{L\mu}j_{L}\,,\quad~~~(\mathcal{O}_{3})_{\{LL\}R}=J^{\mu}_{L}J_{L\mu}j_{R}\,,\\
\nonumber&&(\mathcal{O}_{3})_{\{RR\}L}= J^{\mu}_{R}J_{R\mu}j_{L}\,,\quad~~~(\mathcal{O}_{3})_{\{RR\}R}=J^{\mu}_{R}J_{R\mu}j_{R}\,,\quad~~~(\mathcal{O}_{3})_{\{LR\}L}= J^{\mu}_{L}J_{R\mu}j_{L}\,,\\ \nonumber&&(\mathcal{O}_{3})_{\{LR\}R}=J^{\mu}_{L}J_{R\mu}j_{R}\,,\quad~~~(\mathcal{O}_{4})_{LL}=J_{L}^{\mu}J_{L\mu\nu}j^{\nu}\,,\quad~~~
(\mathcal{O}_{4})_{RR}=J_{R}^{\mu}J_{R\mu\nu}j^{\nu}\,,\\
\nonumber&&(\mathcal{O}_{4})_{LR}=J_{L}^{\mu}J_{R\mu\nu}j^{\nu}\,,\quad~~~(\mathcal{O}_{4})_{RL}=J_{R}^{\mu}J_{L\mu\nu}j^{\nu}\,,\quad~~~ (\mathcal{O}_{5})_{LL}=J_{L}J_{L}^{\mu}j_{\mu}\,,\\
\label{eq:0nubb-LEFT}&&(\mathcal{O}_{5})_{RR}=J_{R}J_{R}^{\mu}j_{\mu}\,,\quad~~~(\mathcal{O}_{5})_{LR}=J_{L}J_{R}^{\mu}j_{\mu}\,,\quad~~~
(\mathcal{O}_{5})_{RL}=J_{R}J_{L}^{\mu}j_{\mu}\,.
\end{eqnarray}
Here $J_{R,L}=\bar{u}_{a}(1\pm\gamma_{5})d_{a}$, $J_{R,L}^{\mu}=\bar{u}_{a}\gamma^{\mu}(1\pm\gamma_{5})d_{a}$ and $
J_{R,L}^{\mu\nu}=\bar{u}_{a}\sigma^{\mu\nu}(1\pm\gamma_{5})d_{a}$ are quark currents with $a=1, 2, 3$ labelling the $SU(3)_{C}$ color indices and $\sigma^{\mu\nu}=\frac{i}{2}[\gamma^{\mu},\gamma^{\nu}]$, the leptonic currents $j_{R,L}=\bar{e}(1\mp\gamma_{5})e^{c}$, $j^{\mu}=\bar{e}\gamma^{\mu}\gamma_{5}e^{c}$ are bilinear combinations of electron fields. It should be noted that the operators $(\mathcal{O}_{2})_{\{LR\}R}$, $(\mathcal{O}_{2})_{\{RL\}L}$ and $(\mathcal{O}_{2})_{\{RL\}R}$ are identically zero due to the identity $J^{\mu\nu}_{L}J_{R\mu\nu}=0$. The most general short-range interaction Lagrangian governing $0\nu\beta\beta$ decay is given by
\begin{equation}
\label{eq:EFLag-SR0nbb}\mathcal{L}_\text{short}=\frac{G^2_F\cos^2\theta_C}{2m_P}\sum_{\chi_1, \chi_2, \chi}\left[\sum^3_{i=1}\epsilon_{i\chi_1\chi_2}^{\chi}(\mathcal{O}_{i})_{\{\chi_1\chi_2\}\chi}+
\sum^{5}_{i=4}\epsilon_{i\chi_1\chi_2}(\mathcal{O}_{i})_{\chi_1\chi_2}\right]+ \mathrm{h.c.}\,,
\end{equation}
where $\chi_1, \chi_2, \chi\in\left\{L, R\right\}$ denote the chiralities of the quark and electron currents, $m_P$ is the proton mass and $\theta_C$ is the Cabibbo angle. Taking into account the phase-space factors (PSFs) and the nuclear matrix elements (NMEs), the inverse half-life of $0\nu\beta\beta$ decay is given by~\cite{Deppisch:2020ztt},
\begin{eqnarray}
\nonumber T_{1/2}^{-1} &=& G_{11+}^{(0)}\left|
\sum_{I=1}^{3}\epsilon_{I}^{L}\mathcal{M}_{I}
+\epsilon_{\nu}\mathcal{M}_{\nu}\right|^{2}+G_{11+}^{(0)}\left|\sum_{I=1}^{3}\epsilon_{I}^{R}\mathcal{M}_{I}\right|^{2}
+G_{66}^{(0)}\left|\sum_{I=4}^{5}\epsilon_{I}\mathcal{M}_{I}\right|^{2}\\
\nonumber&&+2G_{11-}^{(0)}\Re\left[\left(\sum_{I=1}^{3}\epsilon_{I}^{L}\mathcal{M}_{I}
+\epsilon_{\nu}\mathcal{M}_{\nu}\right)\left(\sum_{I=1}^{3}\epsilon_{I}^{R}\mathcal{M}_{I}\right)^{*}\right]\\
\label{eq:inverse-half-life}&&+2G_{16}^{(0)}\Re\left[\left(\sum_{I=1}^{3}\epsilon_{I}^{L}\mathcal{M}_{I}-\sum_{I=1}^{3}\epsilon_{I}^{R}\mathcal{M}_{I}
+\epsilon_{\nu}\mathcal{M}_{\nu}\right)
\left(\sum_{I=4}^{5}\epsilon_{I}\mathcal{M}_{I}\right)^{*}\right]\label{eq:halflife}\,,
\end{eqnarray}
where $\epsilon_{\nu}\equiv m_{\beta\beta}/m_e$, the sum extends over all short-range contributions with different chiral structures, labelled by
$I=(i, \chi_1\chi_2)$ with $i=1, \ldots, 5$ and $\chi_1, \chi_2\in\left\{L, R\right\}$. For $i=1,2,3$, the coefficients $\epsilon_{I}^{L}$ and $\epsilon_{I}^{R}$ correspond to
$\epsilon^{\chi_1\chi_2L}_{i}$ and $\epsilon^{\chi_1\chi_2R}_{i}$, respectively. The quantities $G_{11+}^{(0)}$, $G_{66}^{(0)}$, $G_{11-}^{(0)}$, and $G_{16}^{(0)}$ denote the relevant phase-space factors, while $\mathcal{M}_{\nu}$ and $\mathcal{M}_{I}$ are the nuclear matrix elements for the mass mechanism and the short-range operators. Their numerical values, computed within the interacting boson model, are listed in Ref.~\cite{Deppisch:2020ztt}.

\section{Two-loop integrals of neutrino masses\label{app:2loop-def}}

In the following, we provide the analytical expressions for the two-loop integrals that arise in the calculation of the light neutrino mass in
$0\nu\beta\beta$ decay models. We follow the notation and definitions introduced in Ref.~\cite{vanderBij:1983bw}:
\begin{eqnarray}
&&\left(M_{11},\cdots,M_{1n_1}|M_{21},\cdots,M_{2n_2}|M_{31},\cdots,M_{3n_3}\right)\nonumber\\
&=&\left(\frac{e^{\gamma_E\epsilon}}{i\pi^{d/2}}\right)^2\left(\mu^{4-d}\right)^2\int d^dp\int d^dq\prod_{i=1}^{n_1}\prod_{j=1}^{n_2}\prod_{k=1}^{n_3}\frac{1}{p^2+M_{1i}^2}\frac{1}{q^2+M_{2j}^2}\frac{1}{\left(p+q\right)^2+M_{3k}^2}\label{eq:2loop-int}\,,
\end{eqnarray}
where $d=4-2\epsilon$ is spacetime dimension in dimensional regularization, $\mu$ is an energy scale, and the Wick rotation has been performed. In the case that the loop integral is divergent, one has to carefully evaluate all terms before taking the limit $\epsilon\rightarrow0$. One can use partial fractions to obtain the following relation:
\begin{eqnarray} \left(m,m_0|m_1|m_2\right)&=&\frac{1}{m^2-m_0^2}\big[\left(m_0|m_1|m_2\right)-\left(m|m_1|m_2\right)\big]\,.\label{eq:mm0-mm0}
\end{eqnarray}
It is convenient to express $(m_0|m_1|m_2)$ in terms of the less divergent integral $(2m_0|m_1|m_2)$. The partial $p$ operation of 't Hooft~\cite{tHooft:1972tcz} can give rise to the following relations:
\begin{eqnarray} \left(m_0|m_1|m_2\right)&=&\frac{1}{3-d}\big[m_0^2\left(2m_0|m_1|m_2\right)+m_1^2\left(2m_1|m_0|m_2\right)
+m_2^2\left(2m_2|m_0|m_1\right)\big]\,,\label{eq:m0m1m2-2m0m1m2}
\end{eqnarray}
where $(2m_0|m_1|m_2)$ is a shorthand for $(m_0, m_0|m_1|m_2)$. The analytical expression of $(m_0, m_0|m_1|m_2)$ is given by~\cite{vanderBij:1983bw,Ghinculov:1994sd,McDonald:2003zj},
\begin{eqnarray}
(2m|m_{1}|m_{2})&=&-\frac{1}{2\epsilon^{2}}-\frac{1}{2\epsilon}\left[1-2\log\left(\frac{m^2}{\mu^2}\right)\right]\nonumber\\
&&-\left[\frac{1}{2}+\frac{1}{12}{\pi}^{2}-\log\left(\frac{{m}^{2}}{\mu^{2}}\right)
+\log^{2}\left(\frac{{m}^{2}}{\mu^{2}}\right)+f(a,b)\right]\label{eq:2m-m1-m2-integ}
\end{eqnarray}
where $a=m^2_1/m^2$, $b=m^2_2/m^2$, $\gamma_E\approx0.577$ is the Euler-Mascheroni constant, and the function $f(a, b)$ reads as follows~\cite{McDonald:2003zj},
\begin{eqnarray}
f(a,b)&=& -\frac{1}{2}\log a\log b -\frac{1}{2}\frac{a+b-1}{\sqrt{\phantom{ab}}}\left\{\mathrm{Li_{2}}\left(\frac{-x_{2}}{y_{1}}\right)+
\mathrm{Li_{2}}\left(\frac{-y_{2}}{x_{1}}\right)-\mathrm{Li_{2}}\left(\frac{-x_{1}}{y_{2}}\right)-\mathrm{Li_{2}}\left(\frac{-y_{1}}{x_{2}}\right) \right.\nonumber\\
&&\hspace{3cm}\left.+\mathrm{Li_{2}}\left(\frac{b-a}{x_{2}}\right)+\mathrm{Li_{2}}\left(\frac{a-b}{y_{2}}\right)-
\mathrm{Li_{2}}\left(\frac{b-a}{x_{1}}\right)-\mathrm{Li_{2}}\left(\frac{a-b}{y_{1}}\right)\right\}\,.
\end{eqnarray}
Here we have introduced the parameters
\begin{eqnarray}
x_{1}&=&\frac{1}{2}\left(1+b-a +\sqrt{\phantom{ab}}\right),\qquad
x_{2}=\frac{1}{2}\left(1+b-a -\sqrt{\phantom{ab}}\right),\nonumber\\
y_{1}&=&\frac{1}{2}\left(1+a-b+\sqrt{\phantom{ab}}\right),\qquad y_{2}=\frac{1}{2}\left(1+a-b -\sqrt{\phantom{ab}}\right),\nonumber\\
&&\qquad \sqrt{\phantom{ab}} =\left[1-2(a+b)+(a-b)^{2}\right]^{1/2}\,,
\end{eqnarray}
and $\text{Li}_2(x)$ is the standard di-logarithm function defined by
\begin{eqnarray}
\text{Li}_2(x)=-\int_0^x\frac{\ln(1-y)}{y}dy\,,
\end{eqnarray}
In order to determine the light neutrino mass in our model, we need to evaluate the two-loop integrals $\hat{\mathcal{I}}^{p_1\cdot p_2}$ and $\hat{\mathcal{I}}^{p^2\left(p\cdot q\right)}$ defined as
\begin{eqnarray}
&&\hat{\mathcal{I}}^{p\cdot q}\left(m_a,m_b,m_\alpha,m_\beta,m\right)\nonumber\\
=&&\left(\frac{e^{\gamma_E\epsilon}}{i\pi^{d/2}}\right)^2\left(\mu^{4-d}\right)^2\iint d^dpd^dq\frac{p\cdot q}{\left(p^2-m_a^2\right)\left(p^2-m_b^2\right)\left(q^2-m_\alpha^2\right)\left(q^2-m_\beta^2\right)\left[\left(p+q\right)^2-m^2\right]}\,,\nonumber\\\label{eq:hati-p1p2-def}\nonumber\\
&&\hat{\mathcal{I}}^{p^2\left(p\cdot q\right)}\left(m_a,m_b,m_\alpha,m_\beta,m\right)\nonumber\\
=&&\left(\frac{e^{\gamma_E\epsilon}}{i\pi^{d/2}}\right)^2\left(\mu^{4-d}\right)^2\iint d^dpd^dq\frac{p^2\left(p\cdot q\right)}{\left(p^2-m_a^2\right)\left(p^2-m_b^2\right)\left(q^2-m_\alpha^2\right)\left(q^2-m_\beta^2\right)\left[\left(p+q\right)^2-m^2\right]}\,.\nonumber\\
\end{eqnarray}
Using the identities in Eqs.~(\ref{eq:mm0-mm0}, \ref{eq:m0m1m2-2m0m1m2}, \ref{eq:2m-m1-m2-integ}), we find the complete expression of $\hat{\mathcal{I}}^{p\cdot q}$ and $\hat{\mathcal{I}}^{p^2\left(p\cdot q\right)}$ are
\begin{eqnarray}
\hat{\mathcal{I}}^{p\cdot q}&=&-\frac{1}{2\epsilon}-2+\frac{\left(m_a^2\log\frac{m_a^2}{\mu^2}-m_b^2\log\frac{m_b^2}{\mu^2}\right)\left(m_\alpha^2\log \frac{m_\alpha^2}{\mu^2}-m_\beta^2\log\frac{m_\beta^2}{\mu^2}\right)}{2\left(m_a^2-m_b^2\right)\left(m_\alpha^2-m_\beta^2\right)}\nonumber\\
&&-\frac{\left(m_a^2\log^2\frac{m_a^2}{\mu^2}-m_b^2\log^2\frac{m_b^2}{\mu^2}\right)}{4\left(m_a^2-m_b^2\right)}
-\frac{\left(m_\alpha^2\log^2\frac{m_\alpha^2}{\mu^2}-m_\beta^2\log^2\frac{m_\beta^2}{\mu^2}\right)}{4\left(m_\alpha^2-m_\beta^2\right)}\nonumber\\
&&+\frac{\left(m_a^2\log\frac{m_a^2}{\mu^2}-m_b^2\log\frac{m_b^2}{\mu^2}\right)}{2\left(m_a^2-m_b^2\right)}
+\frac{\left(m_\alpha^2\log\frac{m_\alpha^2}{\mu^2}-m_\beta^2\log\frac{m_\beta^2}{\mu^2}\right)}{2\left(m_\alpha^2-m_\beta^2\right)}\nonumber\\
&&-\frac{\left(m_a^2+m_\alpha^2-m^2\right)}{2\left(m_a^2-m_b^2\right)\left(m_\alpha^2-m_\beta^2\right)}
\Big[m_a^2f\left(\frac{m_\alpha^2}{m_a^2},\frac{m^2}{m_a^2}\right)+m_\alpha^2f\left(\frac{m_a^2}{m_\alpha^2},\frac{m^2}{m_\alpha^2}\right)
+m^2f\left(\frac{m_a^2}{m^2},\frac{m_\alpha^2}{m^2}\right)\Big]\nonumber\\
&&+\frac{\left(m_a^2+m_\beta^2-m^2\right)}{2\left(m_a^2-m_b^2\right)\left(m_\alpha^2-m_\beta^2\right)}
\Big[m_a^2f\left(\frac{m_\beta^2}{m_a^2},\frac{m^2}{m_a^2}\right)+m_\beta^2f\left(\frac{m_a^2}{m_\beta^2},\frac{m^2}{m_\beta^2}\right)
+m^2f\left(\frac{m_a^2}{m^2},\frac{m_\beta^2}{m^2}\right)\Big]\nonumber\\
&&+\frac{\left(m_b^2+m_\alpha^2-m^2\right)}{2\left(m_a^2-m_b^2\right)\left(m_\alpha^2-m_\beta^2\right)}
\Big[m_b^2f\left(\frac{m_\alpha^2}{m_b^2},\frac{m^2}{m_b^2}\right)+m_\alpha^2f\left(\frac{m_b^2}{m_\alpha^2},\frac{m^2}{m_\alpha^2}\right)
+m^2f\left(\frac{m_b^2}{m^2},\frac{m_\alpha^2}{m^2}\right)\Big]\nonumber\\
&&-\frac{\left(m_b^2+m_\beta^2-m^2\right)}{2\left(m_a^2-m_b^2\right)\left(m_\alpha^2-m_\beta^2\right)}
\Big[m_b^2f\left(\frac{m_\beta^2}{m_b^2},\frac{m^2}{m_b^2}\right)+m_\beta^2f\left(\frac{m_b^2}{m_\beta^2},\frac{m^2}{m_\beta^2}\right)
+m^2f\left(\frac{m_b^2}{m^2},\frac{m_\beta^2}{m^2}\right)\Big]\,,\label{eq:hati-p1p2-res}~~~~~~
\end{eqnarray}
and
\begin{eqnarray}
\hat{\mathcal{I}}^{p^2(p\cdot q)}&=&-\frac{m_\alpha^2+m_\beta^2+2m^2}{4\epsilon^2}-\frac{1}{4\epsilon}\Big[2\left(m_a^2+m_b^2\right)+3\left(m_\alpha^2+m_\beta^2\right)
+4\left(1-\log\frac{m^2}{\mu^2}\right)m^2\nonumber\\
&&-2\frac{m_\alpha^4\log\frac{m_\alpha^2}{\mu^2}-m_\beta^4\log\frac{m_\beta^2}{\mu^2}}{m_\alpha^2-m_\beta^2}\Big]-2\left(m_a^2+m_b^2\right)
-\frac{1}{24}\left(42+\pi^2\right)\left(m_\alpha^2+m_\beta^2\right)-\frac{1}{12}\left[\pi^2+18\right]m^2\nonumber\\
&&+\frac{m_a^4\log\frac{m_a^2}{\mu^2}-m_b^4\log\frac{m_b^2}{\mu^2}}{2\left(m_a^2-m_b^2\right)}-\frac{m_a^4\log^2\frac{m_a^2}{\mu^2}
-m_b^4\log^2\frac{m_b^2}{\mu^2}}{4\left(m_a^2-m_b^2\right)}\nonumber\\
&&+\frac{1}{2\left(m_\alpha^2-m_\beta^2\right)}\Big\{\left[m_a^2+m_b^2+3m_\alpha^2-m^2\right]m_\alpha^2\log\frac{m_\alpha^2}{\mu^2}
-\left[m_a^2+m_b^2+3m_\beta^2-m^2\right]m_\beta^2\log\frac{m_\beta^2}{\mu^2}\Big\}\nonumber\\
&&-\frac{\left[\left(m_a^2+m_b^2+2m_\alpha^2-m^2\right)m_\alpha^2\log^2\frac{m_\alpha^2}{\mu^2}
-\left(m_a^2+m_b^2+2m_\beta^2-m^2\right)m_\beta^2\log^2\frac{m_\beta^2}{\mu^2}\right]}{4\left(m_\alpha^2-m_\beta^2\right)}\nonumber\\
&&+\frac{5m^2\log\frac{m^2}{\mu^2}}{2}-\frac{3m^2\log^2\frac{m^2}{\mu^2}}{4}+\frac{\left(m_\alpha^2\log\frac{m_\alpha^2}{\mu^2}
-m_\beta^2\log\frac{m_\beta^2}{\mu^2}\right)\left(m_a^4\log\frac{m_a^2}{\mu^2}
-m_b^4\log\frac{m_b^2}{\mu^2}\right)}{2\left(m_a^2-m_b^2\right)\left(m_\alpha^2-m_\beta^2\right)}\nonumber\\
&&-\frac{\left(m_\alpha^2\log\frac{m_\alpha^2}{\mu^2}-m_\beta^2\log\frac{m_\beta^2}{\mu^2}\right)m^2\log\frac{m^2}{\mu^2}}{2\left(m_\alpha^2-m_\beta^2\right)}\nonumber\\
&&-\frac{m_a^2\left(m_a^2+m_\alpha^2-m^2\right)}{2\left(m_a^2-m_b^2\right)\left(m_\alpha^2-m_\beta^2\right)}
\Big[m_\alpha^2f\left(\frac{m_a^2}{m_\alpha^2},\frac{m^2}{m_\alpha^2}\right)+m^2f\left(\frac{m_a^2}{m^2},\frac{m_\alpha^2}{m^5}\right)
+m_a^2f\left(\frac{m_\alpha^2}{m_a^2},\frac{m^2}{m_a^2}\right)\Big]\nonumber\\
&&+\frac{m_a^2\left(m_a^2+m_\beta^2-m^2\right)}{2\left(m_a^2-m_b^2\right)\left(m_\alpha^2-m_\beta^2\right)}
\Big[m_\beta^2f\left(\frac{m_a^2}{m_\beta^2},\frac{m^2}{m_\beta^2}\right)+m^2f\left(\frac{m_a^2}{m^2},\frac{m_\beta^2}{m^5}\right)
+m_a^2f\left(\frac{m_\beta^2}{m_a^2},\frac{m^2}{m_a^2}\right)\Big]\nonumber\\
&&+\frac{m_b^2\left(m_b^2+m_\alpha^2-m^2\right)}{2\left(m_a^2-m_b^2\right)\left(m_\alpha^2-m_\beta^2\right)}
\Big[m_\alpha^2f\left(\frac{m_b^2}{m_\alpha^2},\frac{m^2}{m_\alpha^2}\right)+m^2f\left(\frac{m_b^2}{m^2},\frac{m_\alpha^2}{m^5}\right)
+m_b^2f\left(\frac{m_\alpha^2}{m_b^2},\frac{m^2}{m_b^2}\right)\Big]\nonumber\\
&&-\frac{m_b^2\left(m_b^2+m_\beta^2-m^2\right)}{2\left(m_a^2-m_b^2\right)\left(m_\alpha^2-m_\beta^2\right)}
\Big[m_\beta^2f\left(\frac{m_b^2}{m_\beta^2},\frac{m^2}{m_\beta^2}\right)+m^2f\left(\frac{m_b^2}{m^2},\frac{m_\beta^2}{m^5}\right)
+m_b^2f\left(\frac{m_\beta^2}{m_b^2},\frac{m^2}{m_b^2}\right)\Big]\,.
\end{eqnarray}

\section{\label{app:add-onbb-models}Some typical $0\nu\beta\beta$ models for the short-range dim-11 effective operators}

A large class of UV completions of the dim-11 $0\nu\beta\beta$ decay operators has been identified~\cite{Li:2026dim11NDBDsup}. In this appendix, we present several additional simple models beyond the representative example discussed in section~\ref{sec:representative-model}. We focus on scenarios containing at most one color-triplet mediator, with all remaining mediators being colorless.

\subsection{\label{app:2nd-colorless-model}A second $0\nu\beta\beta$ model with colorless mediators}

\begin{figure}[htbp]
\centering
\includegraphics[scale=0.65]{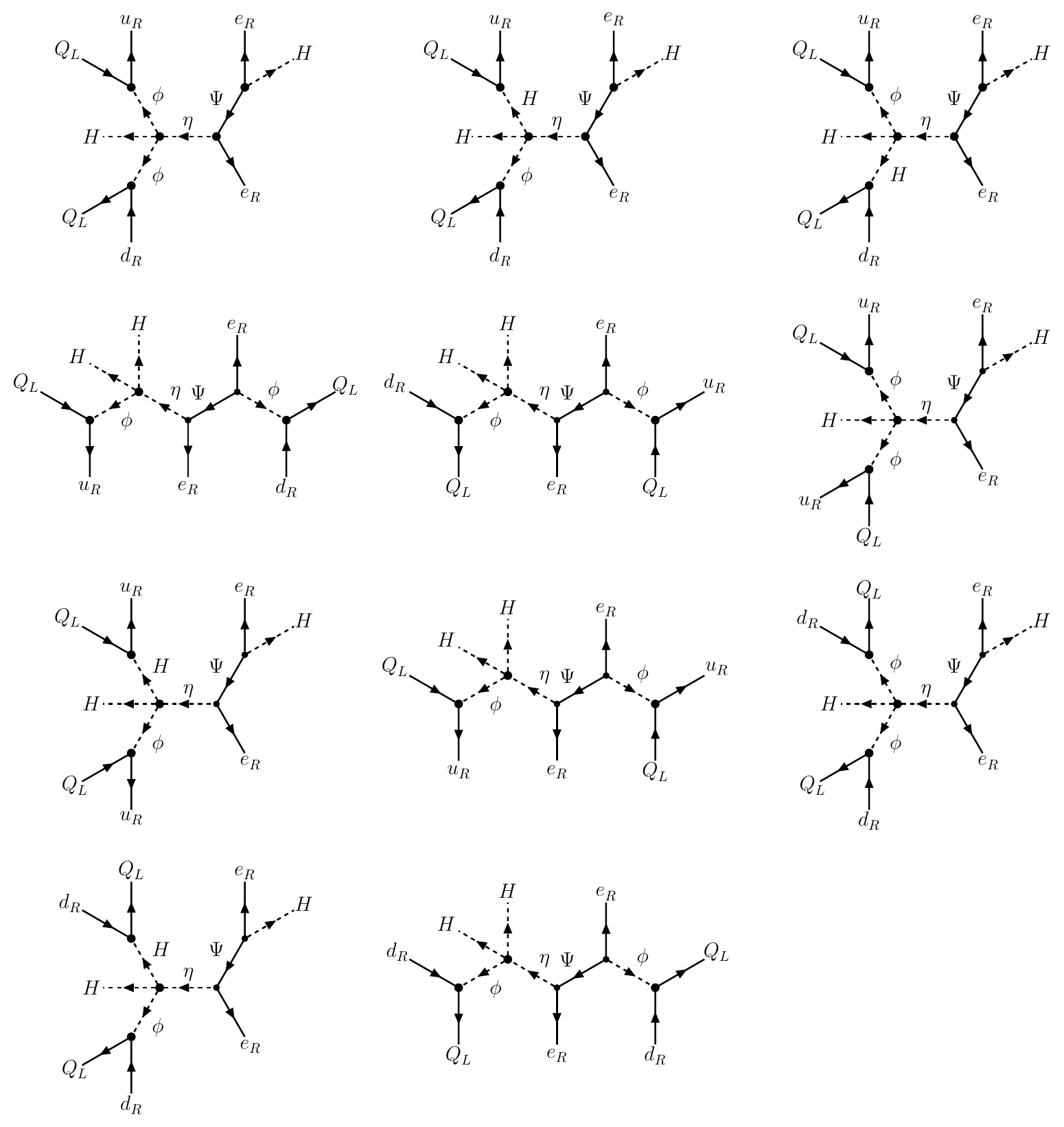}
\caption{\label{fig:model-su3sing-2s1f} The diagrams for the $0\nu\beta\beta$ decay process in the second $0\nu\beta\beta$ model with colorless mediators, where two colorless scalar doublets $\phi$, $\eta$ and one colorless fermion doublet $\Psi$ are introduced. The first five diagrams represent UV completions of the dim-11 $0\nu\beta\beta$ operators $\mathcal{O}_{1a, 1b}$, the subsequent three diagrams are UV completions of $\mathcal{O}_{2a, 2b}$, and  the last three diagrams realize the effective operators
$\mathcal{O}_{3a, 3b}$. }
\end{figure}

Similar to the representation model studied in section~\ref{sec:representative-model}, this model again contains only three colorless new fields, consisting of two distinct scalars and one vector-like fermion. Their transformation properties under the SM gauge group are
\begin{eqnarray}
\label{eq:field-contents-model2}\phi\equiv\begin{pmatrix}
\phi^+\\\phi^0 \end{pmatrix}\sim(\bm{1}, \bm{2}, 1/2, S)\,,~~\eta\equiv\begin{pmatrix}\eta^{++}\\\eta^+
\end{pmatrix}\sim(\bm{1}, \bm{2}, 3/2, S)\,,~~\Psi=\begin{pmatrix}
\Psi^{+}\\ \Psi^0 \end{pmatrix}\sim(\bm{1}, \bm{2}, 1/2, F)\,.~~~
\end{eqnarray}
All these three fields transform as $SU(2)_L$ doublets. In particular, the scalar doublets $\phi$ and $\eta$ are identical to those appearing in the representative model. The new fermion $\Psi$ has the same SM gauge charges as the charge conjugate of the left-handed lepton doublet. However, unlike the SM leptons, it is a vector-like fermion. In addition to the SM interaction Lagrangian, the Yukawa interactions invariant under the SM gauge symmetry can be written as
\begin{eqnarray}
-\mathcal{L}&=&-\mathcal{L}_{\text{SM}}+y_1\,\overline{u_R}\phi^Ti\sigma_2Q_L+y_2\,\overline{Q_L}\phi d_R+y_3\,\overline{\Psi}\widetilde{H}e_R^c+y_4\,\overline{e_R}\eta^\dagger\Psi+\text{h.c.}\,,\label{eq:Yukawa-model2}
\end{eqnarray}
where the flavor indices associated with the Yukawa couplings $y_1$, $y_2$, $y_3$ and $y_4$ are omitted for simplicity. For the $0\nu\beta\beta$ decay processes under consideration, only the interactions involving the first-generation quarks and leptons are relevant. The scalar potential $V(H, \eta, \phi)$ describing the interactions among the SM Higgs doublet $H$ and the additional scalar fields $\eta$ and $\phi$ takes the following form,
\begin{eqnarray}
V(H,\eta,\phi)&=&-\mu^2H^{\dagger}H+m_\phi^2\phi^\dagger\phi+m_\eta\eta^\dagger\eta+\Big(m^2\phi^\dagger H+\text{h.c.}\Big)+\lambda\left(H^{\dagger}H\right)^2+\lambda_1\left(\eta^\dagger\eta\right)^2\nonumber\\
&&+\lambda_2\left(\eta^\dagger\eta\right)\left(\phi^{\dagger}\phi\right)+\lambda_3\left(\eta^\dagger\phi\right)
\left(\phi^{\dagger}\eta\right)+\lambda_4\left(\eta^{\dagger}\eta\right)\left(H^{\dagger}H\right)+\lambda_5\left(\eta^{\dagger} H\right)\left(H^{\dagger}\eta\right)\nonumber\\
&&+\lambda_6\left(\phi^{\dagger}\phi\right)\left(\phi^{\dagger}\phi\right)+\lambda_7\left(\phi^{\dagger}\phi\right)\left(H^{\dagger}H\right)
+\lambda_8\left(\phi^{\dagger}H\right)\left(H^{\dagger}\phi\right)+\Big[\lambda_9\left(\eta^{\dagger}\eta\right)\left(\phi^{\dagger}H\right)\nonumber\\
&&+\lambda_{10}\left(\eta^{\dagger}H\right)\left(\phi^{\dagger}\eta\right)+\lambda_{11}\left(H^{\dagger}\widetilde{\phi}\right)\left(H^{\dagger}\eta\right)
+\lambda_{12}\left(H^{\dagger}\widetilde{\phi}\right)\left(\phi^{\dagger}\eta\right)+\lambda_{13}\left(\phi^{\dagger}H\right)^2\nonumber\\
\label{eq:potential-model2}&&+\lambda_{14}\left(\phi^{\dagger}\phi\right)\left(\phi^{\dagger}H\right)+\lambda_{15}\left(\phi^{\dagger}H\right)\left(H^{\dagger}H\right)+\text{h.c.}\Big]\,.
\end{eqnarray}
Given the interactions in Eqs.~(\ref{eq:potential-model2},\ref{eq:Yukawa-model2}), the $0\nu\beta\beta$ decay can be mediated by the diagrams shown in figure~\ref{fig:model-su3sing-2s1f}. It is evident that either the interaction vertex $\lambda^{*}_{11}\left(\widetilde{\phi}^{\dagger}H\right)\left(\eta^{\dagger}H\right)$ or $\lambda_{12}\left(H^{\dagger}\widetilde{\phi}\right)\left(\phi^{\dagger}\eta\right)$ is present in all diagrams in figure~\ref{fig:model-su3sing-2s1f}. This vertex would vanish if $\phi$ were identified with the SM Higgs doublet
$H$. Therefore, $\phi$ must be a scalar doublet distinct from $H$, even though they carry identical SM gauge quantum numbers.

\subsection{\label{app:models-color3}$0\nu\beta\beta$ models with two new $SU(3)_C$ singlets and one new $SU(3)_C$ triplet }

We discuss the $0\nu\beta\beta$ decay models that involve colored fields in the following. We restrict ourselves to the simple case in which only one color-triplet field beyond the SM is present. Under this assumption, there exist only eight such models, which are summarized in table~\ref{ta:model-su3-2s1t}. Each of these models contains two colorless fields and one color-triplet field. Apart from the fields $\zeta$, $\eta$, $\Psi$ introduced in Eq.~\eqref{eq:scalars-model1} and Eq.~\eqref{eq:field-contents-model2}, the additional field multiplets are
\begin{eqnarray}
\nonumber&&\chi\sim (\bm{1}, \bm{1}, 1, S)\,,\quad \Gamma\sim (\bm{3}, \bm{1}, -1/3, F)\,,\quad \Theta\sim (\bm{3}, \bm{1}, 2/3, F)\,,\\
\label{eq:fields-model3c}&&\Omega\sim (\bm{3}, \bm{2}, -5/6, F)\,,\quad  \Delta\sim (\bm{3}, \bm{2}, 7/6, F)\,,
\end{eqnarray}
where all fermionic fields transform as color triplets, while the scalar field is colorless. In all these models, the light neutrino masses are generated at the two-loop level. The models MF-3i-1, MF-3i-2, MF-3i-3, MF-3i-4, MF-3i-6, and MF-3i-7 contain two scalar fields and one fermionic field, whereas the remaining models MF-3i-40 and MF-3i-41 involve one scalar field and two fermionic fields. Remarkably, in the models MF-3i-1, MF-3i-2, MF-3i-3, and MF-3i-4 which are the UV completions of $\mathcal{O}_{2a, 2b}$ or $\mathcal{O}_{3a, 3b}$, there exists only a single Feynman diagram contributing to the $0\nu\beta\beta$ decay process. In all these four cases, the Feynman diagram is based on the topology T4-6 and contains five propagators. Furthermore, all propagators appearing in these diagrams correspond to new fields introduced in Eqs.~(\ref{eq:scalars-model1}, \ref{eq:field-contents-model2}, \ref{eq:fields-model3c}).

\begin{table}[t!]
\centering
\belowbottomsep=-0.8ex
\resizebox{\linewidth}{!}{
\begin{tabular}{|cclc|} \hline\hline
No.&Fields&\multicolumn{1}{c}{$\mathcal{O}_i$: Model-diagram}&\multicolumn{1}{c|}{Relevant  interactions}\\ \hline
			MF-3i-1&$\chi,\,\zeta,\,\Gamma$&\multicolumn{1}{c}{$\mathcal{O}_{2a},\mathcal{O}_{2b}$: T4-6-102}&\makecell[c]{$y_{1}\,\overline{\Gamma}H^\dagger Q_L+y_{2}\,\overline{e_R}e_R^c\zeta$\\
$+y_{3}\,\overline{u_R}\Gamma\chi+\mu_{1}\,\chi^*\chi\zeta$}\\ \hline
			MF-3i-2&$\chi,\,\zeta,\,\Theta$&\multicolumn{1}{c}{$\mathcal{O}_{3a},\mathcal{O}_{3b}$: T4-6-119}&\makecell[c]{$y_{1}\,\overline{Q_L}\widetilde{H} \Theta+y_{2}\,\overline{e_R}e_R^c\zeta^c$\\			$+y_{3}\,\overline{\Theta}d_R\chi+\mu_{1}\,\chi^*\chi^*\zeta$}\\\hline

MF-3i-3&$\chi,\,\zeta,\, \Omega$&\multicolumn{1}{c}{$\mathcal{O}_{3a},\mathcal{O}_{3b}$: T4-6-122}&\makecell[c]{$y_{1}\,\overline{\Omega}\widetilde{H} d_R+y_{2}\,\overline{e_R}e_R^c\zeta^*$\\
$+y_{3}\,\overline{Q_L} \Omega\chi+\mu_{1}\,\chi^*\chi^*\zeta$}\\\hline
			MF-3i-4&$\chi,\,\zeta,\,\Delta$&\multicolumn{1}{c}{$\mathcal{O}_{2a},\mathcal{O}_{2b}$: T4-6-99}&\makecell[c]{$y_{1}\,\overline{u_R}H^\dagger \Delta+y_{2}\,\overline{e_R}e_R^c\zeta^*$\\
$+y_{3}\,\overline{\Delta}Q_L \chi+\mu_{1}\, \chi^* \chi^*\zeta$}\\ \hline

MF-3i-6&$\zeta,\eta,\Omega$&\begin{tabular}{cc}
$\mathcal{O}_{1a},\mathcal{O}_{1b}$:&\makecell[l]{T4-2-15, T4-6-42, T4-6-56, T5-2-15,\\
T5-7-1, T6-1-5, T7-5-90, T7-5-97}\\\hline
$\mathcal{O}_{2a},\mathcal{O}_{2b}$:&T5-2-44, T5-7-18\\\hline
$\mathcal{O}_{3a},\mathcal{O}_{3b}$:&T4-2-75, T6-1-23
\end{tabular}
&\makecell[c]{$y_{1}\,\overline{ \Omega}\widetilde{H} d_R+y_{2}\,\overline{u_R} \eta^Ti\sigma_2\Omega$\\
$+y_{3}\,\overline{e_R}e_R^c\zeta^*+\mu_{1}\,\zeta^*H^Ti\sigma_2\eta$}\\\hline

MF-3i-7&$\zeta,\eta,\Delta$&\begin{tabular}{cc} $\mathcal{O}_{1a},\mathcal{O}_{1b}$:&\makecell[l]{T4-2-3, T4-6-35, T4-6-49, T5-2-23,\\
T5-7-11, T6-1-1, T7-5-112, T7-5-120}\\\hline
$\mathcal{O}_{2a},\mathcal{O}_{2b}$:&T4-2-50, T6-1-9\\\hline
$\mathcal{O}_{3a},\mathcal{O}_{3b}$:&T5-2-64, T5-7-28
\end{tabular} &\makecell[c]{$y_{1}\,\overline{\Delta}\eta d_R+y_{2}\,\overline{e_R}e_R^c\zeta^*$\\
$+y_{3}\,\overline{u_R} H^\dagger \Delta+\mu_{1}\,\zeta^*H^Ti\sigma_2\eta$}\\\hline

MF-3i-40&$\eta,\Psi,\Omega$&\begin{tabular}{cc}
$\mathcal{O}_{1a},\mathcal{O}_{1b}$:&\makecell[c]{T4-4-71, T5-4-5, T5-5-9, T7-7-129,\\
T7-7-148, T7-7-243, T7-7-257, T7-9-3}\\\hline
$\mathcal{O}_{2a},\mathcal{O}_{2b}$:&T5-4-178, T7-9-131\\\hline
$\mathcal{O}_{3a},\mathcal{O}_{3b}$:&T4-4-288, T5-5-136
\end{tabular} &\makecell[c]{
$y_{1}\,\overline{e_R}\eta^Ti\sigma_2\Psi+y_{2}\,\overline{\Omega}\widetilde{H}d_R$\\
$+y_{3}\,\overline{u_R}\eta^Ti\sigma_2\Omega+y_{4}\,\overline{\Psi}\widetilde{H}e_R^C$}\\\hline

MF-3i-41&$\eta,\Psi,\Delta$&\begin{tabular}{cc}$\mathcal{O}_{1a},\mathcal{O}_{1b}$:&\makecell[c]{T4-4-13, T5-4-53, T5-5-1, T7-7-170,\\
T7-7-177, T7-7-237, T7-7-250, T7-9-23}\\\hline
$\mathcal{O}_{2a},\mathcal{O}_{2b}$:&T4-4-223, T5-5-87\\\hline
$\mathcal{O}_{3a},\mathcal{O}_{3b}$:&T5-4-236, T7-9-171
\end{tabular} &\makecell[c]{ $y_{1}\,\overline{e_R}H^\dagger\Psi^C+y_{2}\,\overline{\Delta}\eta d_R$\\
$+y_{3}\,\overline{u_R} H^\dagger \Delta+y_{4}\,\overline{e_R}\eta^\dagger\Psi$}\\ \hline\hline
\end{tabular}}
\caption{\label{ta:model-su3-2s1t} Summary of the $0\nu\beta\beta$ decay models with two colorless fields and one color-triplet mediator. }
\end{table}

The remaining four models, MF-3i-6, MF-3i-7, MF-3i-40 and MF-3i-41, provide UV completions of the effective operators $\mathcal{O}_{1a, 1b}$, $\mathcal{O}_{2a, 2b}$ and $\mathcal{O}_{3a, 3b}$. For each of these models, there are twelve distinct Feynman diagrams contributing to the
$0\nu\beta\beta$ decay process. All of these diagrams involve five mediators, comprising three new fields as well as the SM Higgs field and right-handed quarks. Following the same strategy as in section~\ref{sec:representative-model}, the predictions for the $0\nu\beta\beta$ decay half-life and the light neutrino masses can be obtained. However, the corresponding calculations are technically involved and lengthy, and are therefore deferred to future work.


\providecommand{\href}[2]{#2}\begingroup\raggedright\endgroup

\end{document}